\documentclass[runningheads]{llncs}
\usepackage[T1]{fontenc}
\usepackage{graphicx}
\usepackage{amsmath}   
\usepackage{amssymb}   
\usepackage{bm}   
\usepackage{booktabs}
\usepackage{siunitx}
\usepackage{multirow}
\usepackage{algorithm}
\usepackage{algpseudocode}
\usepackage{xcolor}
\usepackage{colortbl}
\usepackage{adjustbox}
\usepackage{makecell}
\usepackage{xeCJK}
\newcommand{\mymethod}{TaCCS-DFA}

\makeatletter
\renewcommand\subsection{\@startsection{subsection}{2}{\z@}%
  {-10\p@ \@plus -2\p@ \@minus -2\p@}%
  {4\p@ \@plus 1\p@ \@minus 1\p@}%
  {\normalfont\normalsize\bfseries}}
\makeatother
\begin{document}
\title{Focus on What Matters: Fisher-Guided Adaptive Multimodal Fusion for Vulnerability Detection}
\titlerunning{Fisher-Guided Adaptive Multimodal Fusion for Vulnerability Detection}
%
%

\author{Yun Bian\inst{1}\thanks{All authors are affiliated with University of Chinese Academy of Sciences, Beijing, China.} \and
Yi Chen\inst{1} \and
HaiQuan Wang\inst{1} \and
Shihao Li\inst{1} \and
Zhe Cui\inst{1}\thanks{Corresponding author.}}
\authorrunning{Y. Bian et al.}
%
\institute{Chengdu Institute of Computer Applications, Chinese Academy of Sciences,
Chengdu, China\\
\email{\{bianyun, cuizhe\}@casit.com.cn}\\
\email{\{chenyi24, wanghaiquan22, lishihao25\}@mails.ucas.ac.cn}}
\maketitle              
\begin{abstract}
False negatives in software vulnerability detection can lead vulnerable functions to directly enter the build and release pipeline, thereby triggering severe security hazards; to improve the comprehensiveness and accuracy of detection, multimodal vulnerability detection techniques that integrate code sequences and graph structures have gradually become the current mainstream paradigm. Existing methods typically jointly model source code sequences and 
code property graph, but simple concatenation or general interaction mechanisms struggle to fully extract control-flow and data-dependency information in the graph modality that is relevant to vulnerability discrimination, resulting in limited performance gains under real data distributions. To address this issue, this paper proposes the TaCCS-DFA framework, which quantifies the sensitivity of feature directions to classification decisions using Fisher information, constrains cross-modal attention within task-sensitive structural subspaces, and dynamically regulates fusion strength on a per-sample basis through adaptive gating, thereby suppressing the propagation of redundancy and noise while preserving the advantages of semantic representations. Theoretical analysis indicates that, under the isotropic perturbation assumption, this mechanism can tighten the upper bound of the output error from $O(\varepsilon)$ to $O(\sqrt{k/d}\cdot\varepsilon)$. Experiments on the four real-world benchmarks BigVul, PrimeVul, Devign, and ReVeal demonstrate that TaCCS-DFA, with only a 3.4\% increase in inference latency, improves the F1 score by up to 6.3 percentage points and achieves a superior balance between precision and recall in highly imbalanced scenarios. The above results indicate that selectively extracting task-relevant control flow and data dependency information from CPG helps alleviate false-negative issues in practical vulnerability detection scenarios and holds application value for DevSecOps pipelines.
\end{abstract}
\section{Introduction}
\label{sec:Introduction}
Software vulnerabilities constitute the core risk source threatening the security of modern software supply chains. The outbreaks of high-severity vulnerabilities such as Log4Shell (CVE-2021-44228) and Heartbleed (CVE-2014-0160)\cite{nvd2021cve44228,csrb2022log4j,nvd2014cve0160,durumeric2014heartbleed} demonstrate that security flaws in a single component can rapidly propagate to millions of downstream systems through dependency relationships\cite{ruan2025propagationbasedvulnerabilityimpactassessment}, resulting in inestimable security threats and economic losses. In real-world software repositories, the proportion of vulnerable functions is typically less than 5\%; 
once a detection system produces false negatives, the associated vulnerable functions may continue to propagate along the build, test, and release pipelines, forming security vulnerabilities that are difficult to trace. Although traditional manual code audits achieve high accuracy, they can no longer keep pace with the ever-growing code scale and iteration speed; statistics show that security experts can review only approximately 150 lines of code per hour on average\cite{mcgraw2006software}. Automated vulnerability detection is therefore a key component of DevSecOps for identifying defects before code submission or release.

In the practice of manual code review, security experts typically employ a dual-track cognitive strategy\cite{munger2005poor}: 
on one hand, they examine the source code text to grasp the program's semantic logic; on the other hand, for potential risk points such as pointer dereferences and memory allocations, experts mentally simulate the program's execution paths, tracking data transfers between variables and state transitions. This cognitive process is conceptually equivalent to performing directed path traversals in the topological structure of the Code Property Graph (CPG)\cite{yamaguchi2014modeling}. This method combining semantic understanding with structural analysis is particularly effective for identifying logical vulnerabilities such as use-after-free (UAF) or buffer overflows, which often cannot be reliably detected solely through local lexical patterns and instead require inference based on cross-statement structural relationships among variable definitions, resource releases, and subsequent accesses. Although pretrained language/code models have demonstrated strong capabilities in code understanding, they essentially apply attention to linear token sequences and cannot explicitly model topological structures of nonlinear sequences such as control-flow dependencies and cross-statement data propagation paths; such structures precisely constitute the core discriminative signals for the aforementioned logical vulnerabilities, indicating the irreplaceable role of graph structure modeling in vulnerability detection tasks. Recent multimodal methods therefore combine pretrained models for Natural Code Sequence (NCS) with graph neural networks over CPG to improve vulnerability detection\cite{zhou2019devign,chakraborty2020deep,liu2025vul}.

Most studies adopt simple feature concatenation, linear interpolation, or generic cross-attention mechanisms to aggregate NCS and CPG representations. This approach implicitly assumes an unstable premise: that introducing an additional modality will necessarily bring effective information gains. However, in current code multimodal tasks, fusion mechanisms encounter significant bottlenecks in leveraging the complementary value of CPG, whose root causes lie in two mutually reinforcing fundamental problems. On one hand, pretrained models such as CodeBERT and CodeT5\cite{feng2020codebert,wang2021codet5} have already implicitly encoded a substantial amount of syntactic and shallow structural information from large-scale code corpora, resulting in significant information redundancy and subspace overlap between NCS and CPG representations; consequently, naive fusion introduces a large volume of redundant signals. On the other hand, a clear gap remains between the feature extraction capabilities of existing graph neural networks for CPG and the modeling capabilities of pretrained language models for NCS. Structural signals in CPG have not yet been fully transformed into stable discriminative representations; when fused, they mix with noise, thereby diluting the discriminative signals of NCS. The limited fusion effectiveness exerts particularly prominent impacts at the engineering level: in highly class-imbalanced real-world code repositories, even a minor decline in recall rate means that a large number of vulnerable functions are falsely negatively detected and flow into production environments, with potential harm far exceeding the manual review burden caused by false positives. Therefore, achieving high-recall and low-false-positive vulnerability detection under the latency constraints of DevSecOps pipelines constitutes the core challenge in advancing multimodal fusion toward practical engineering deployment.

To address the aforementioned issues, this paper introduces the Fisher Information Matrix (FIM) as the metric for task relevance. Unlike traditional attention mechanisms that rely on local similarity between features, Fisher information can directly quantify the sensitivity of feature perturbations to classification decisions\cite{karakida2019universal,amari2019fisher}, thereby identifying feature subspaces that make substantial contributions to the task objective. Based on this, the paper proposes the TaCCS-DFA (Task-Conditioned Complementary Subspace with Dynamic Fisher Attention) framework: it estimates the Fisher principal subspace through an efficient online low-rank approximation algorithm\cite{oja1982simplified,liang2023optimality}, constrains cross-modal attention to task-sensitive directions, and selectively extracts structural features complementary to NCS from CPG representations; simultaneously, it introduces an adaptive gating mechanism that dynamically regulates the fusion ratio according to each sample's structural complexity, realizing sample-level multimodal adaptive enhancement. This task-oriented selective fusion strategy not only prevents the propagation of redundant information between modalities but also alleviates noise interference arising from asymmetric feature extraction capabilities, thereby significantly improving detection performance while maintaining computational efficiency.

The main contributions of this paper are as follows:

(1) \textbf{Problem Analysis and Fusion Paradigm}: Through experimental analysis, this paper reveals two key challenges faced by multimodal vulnerability detection in utilizing structural evidence under real-world imbalanced scenarios, namely feature redundancy and asymmetry in feature extraction capabilities. To address these issues, this paper proposes a task-conditioned complementary fusion strategy, which uses Fisher information as the metric for feature task-relevance and transforms cross-modal interaction from full-spectrum matching based on content similarity to subspace-selective fusion based on task sensitivity.

(2) \textbf{Theoretical Analysis}: This paper analyzes the robustness of the Fisher-guided attention mechanism from the perspective of information geometry. We prove that, under the isotropic perturbation assumption, by constraining the attention within the $k$-dimensional Fisher principal subspace, the upper bound of the output error can be tightened from $O(\varepsilon)$ to $O(\sqrt{k/d}\cdot\varepsilon)$, where $d$ denotes the total feature dimension and $k$ is the dimension of the Fisher principal subspace with $k \ll d$. This result provides a theoretical foundation for understanding the noise suppression effect of task-oriented feature selection.

(3) \textbf{Method Design and Experimental Validation}: This paper proposes the TaCCS-DFA framework. By employing online low-rank Fisher subspace estimation and adaptive gating mechanisms, the framework achieves effective task-oriented fusion under a low computational overhead, with inference latency increasing by only 3.4\%, demonstrating strong practical deployment value. Experiments conducted on four benchmark datasets, BigVul, Devign, ReVeal, and PrimeVul\cite{fan2020ac,zhou2019devign,chakraborty2020deep,ding2025vulnerability}, show that the framework consistently delivers stable performance improvements across multiple backbone networks and achieves state-of-the-art results. In particular, on the highly class-imbalanced BigVul dataset, the F1 score reaches 87.80\%, representing a 6.3 percentage point improvement over the existing best-performing method, while maintaining low calibration error.

\section{Background and Motivation}
This chapter defines the multimodal representations of code and the geometric properties of Fisher information, and reveals the modal redundancy and asymmetry issues in existing fusion paradigms through experimental validation analysis.

\subsection{Preliminaries}
\begin{figure}[t]
    \centering
    \includegraphics[width=0.9\linewidth]{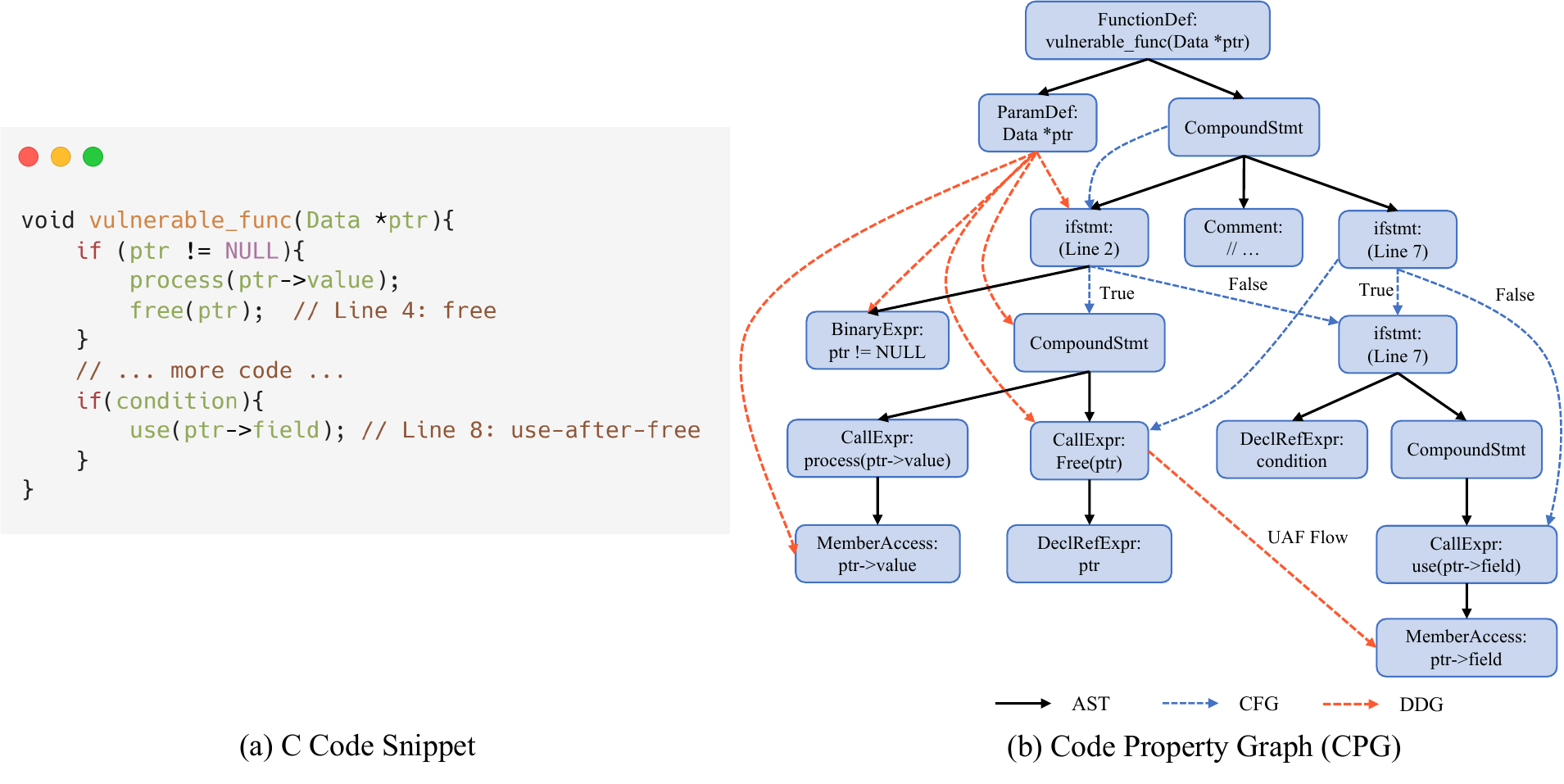}
    \caption{Code and CPG example. (a) A C code snippet containing a UAF vulnerability; (b) the corresponding CPG.}
    \label{fig:code_cpg_example}
\end{figure}

\textbf{Multimodal Representations of Code.}
Source code can be modeled as two complementary modalities: NCS extracts embeddings $\mathbf{H}_{\text{ncs}} \in \mathbb{R}^{L \times d}$ via pretrained language models; CPG is modeled as a directed graph $\mathcal{G} = (\mathcal{V}, \mathcal{E})$, where edges encode control-flow (CFG) and data-dependency (DDG) relationships, and representations $\mathbf{H}_{\text{cpg}} \in \mathbb{R}^{|\mathcal{V}| \times d}$ are obtained through graph neural networks. Figure~\ref{fig:code_cpg_example} illustrates a UAF vulnerability and its CPG structure.

\textbf{Fisher Information as Task-Relevance Metric.}
The FIM quantifies the sensitivity of feature perturbations to classification decisions and is defined as:
\begin{equation}
\mathbf{F}(\mathbf{z}) = \mathbb{E}_{x,y \sim p_{\text{data}}} \left[ \nabla_{\mathbf{z}} \log p_{\theta}(y|x) \cdot \nabla_{\mathbf{z}} \log p_{\theta}(y|x)^\top \right].
\label{eq:fim_def}
\end{equation}
Feature directions with high Fisher information correspond to the most sensitive regions of the decision boundary. This paper utilizes the Fisher principal subspace to guide the attention mechanism, selectively extracting task-relevant structural features from CPG representations.
\subsection{Motivations}
Existing fusion paradigms implicitly assume that additional modalities necessarily bring effective information gains, but analysis on the BigVul dataset indicates that this assumption does not hold.
\begin{figure}[t]
    \centering
    \includegraphics[width=\linewidth]{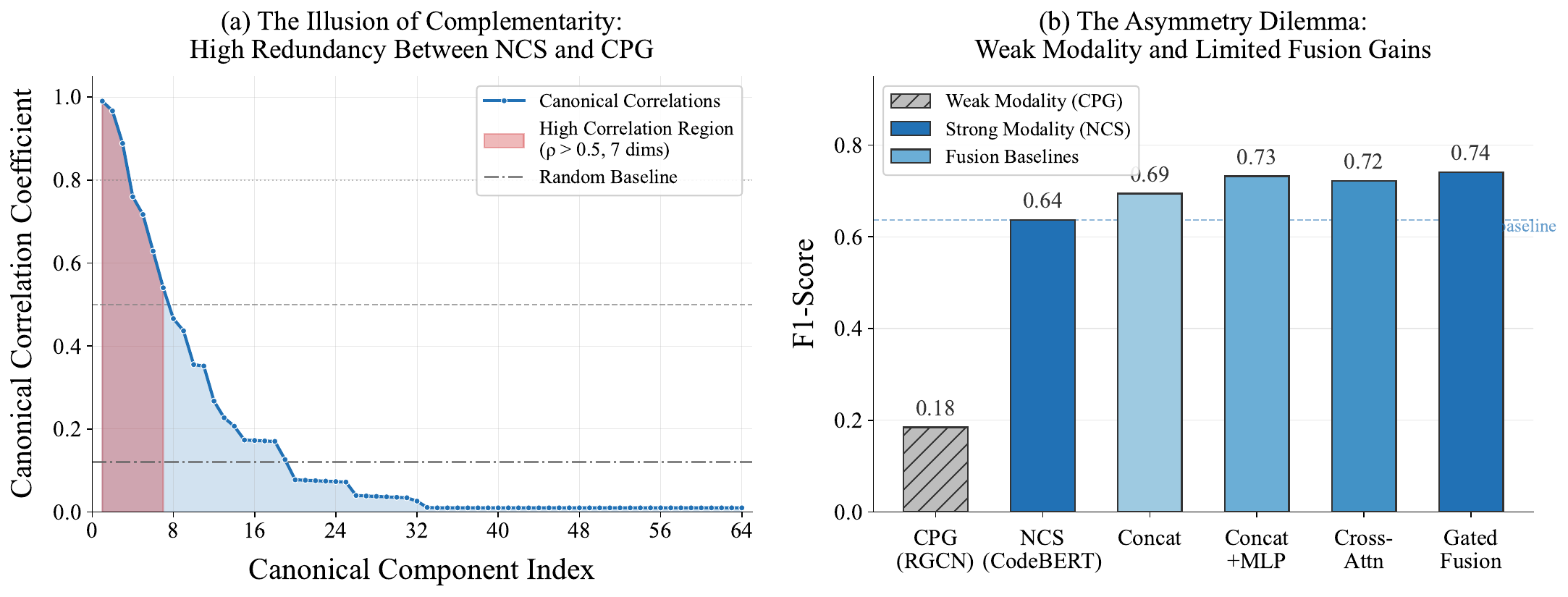}
    \vspace{-0.5cm}
    \caption{Feature space analysis. (a) The CKA similarity between NCS and CPG representations reaches 0.68; (b) Comparison of single-modal detection performance.}
    \label{fig:motivation_analysis}
\end{figure}

\textbf{Information Redundancy.}
Modern pretrained code models have already implicitly encoded a considerable degree of syntactic and control-flow structural information during their self-supervised learning process on large-scale corpora. This implies that a substantial portion of the explicit structural knowledge carried by CPG overlaps with NCS representations. To quantify this phenomenon, this paper computes the centered kernel alignment similarity between NCS features extracted by CodeBERT and CPG features extracted by the Relational Graph Convolutional Network (RGCN)\cite{pmlr-v97-kornblith19a}. As shown in Figure~\ref{fig:motivation_analysis}(a), the linear CKA score between the two reaches 0.68, far higher than the typical near-zero value between randomly initialized features. This result indicates that the feature manifolds of NCS and CPG exhibit significant subspace overlap. Blind feature concatenation or attention-based fusion---i.e., full-spectrum matching---forces the model to process a large volume of redundant signals, which not only increases the computational burden but may also cause the model to converge to a suboptimal solution in an overparameterized space.

\textbf{Feature Extraction Asymmetry.}
In addition to information redundancy, there exists a significant gap between the feature extraction capability of existing graph neural networks for CPG and the modeling capability of pretrained language models for NCS. As shown in Figure~\ref{fig:motivation_analysis}(b), the RGCN model relying solely on CPG achieves an F1 score of less than 0.20, while CodeBERT relying on NCS exceeds 0.63. This gap reflects that current graph encoders still face bottlenecks in feature extraction on complex program graphs and also indicates that the structural information in CPG has not yet been fully transformed into stable discriminative representations. This modal asymmetry poses challenges to the fusion mechanism: when NCS representations are indiscriminately fused with insufficiently extracted CPG representations, the redundant and noisy features mixed in the latter dilute the discriminative signals of the former. Simple concatenation fusion brings only a 5.8 percentage point improvement in F1 score compared to using CodeBERT alone, further illustrating that existing fusion mechanisms still insufficiently utilize the complementary information of CPG. The existing performance bottleneck is closer to the lack of effective distinction between task-relevant structural signals and noise components, rather than the CPG modality itself lacking discriminative value. Comparisons with traditional static analyzers are provided in Appendix~\ref{appendix:static_analysis}.

\textbf{Task-Conditioned Feature Selection.}
Combining the modal features of NCS and CPG with the above analysis, the core contradiction of code multimodal fusion can be stated as follows: CPG contains essential structural information for complex vulnerability detection, yet existing modal fusion methods struggle to effectively extract this information, resulting in feature dilution when fused with NCS and causing existing fusion mechanisms to face significant bottlenecks in using its complementary value. The key to resolving this problem lies in establishing a task-oriented feature selection mechanism that precisely identifies and retains structural subspaces in CPG representations that make substantial contributions to the binary classification task of vulnerability detection, while suppressing the propagation of redundant and low-information signals.

Based on this idea, this paper introduces Fisher information as the metric for task relevance. Unlike traditional attention mechanisms that rely on local similarity between features, Fisher information characterizes the sensitivity of feature perturbations to classification loss and can directly quantify the impact of specific feature directions on the task decision boundary. With the aid of this geometric tool, the proposed method can dynamically locate high-sensitivity structural feature subsets in CPG representations, thereby achieving selective enhancement of complementary information.

\section{Methodology}
\begin{figure*}[t]
\centering
\includegraphics[width=\linewidth]{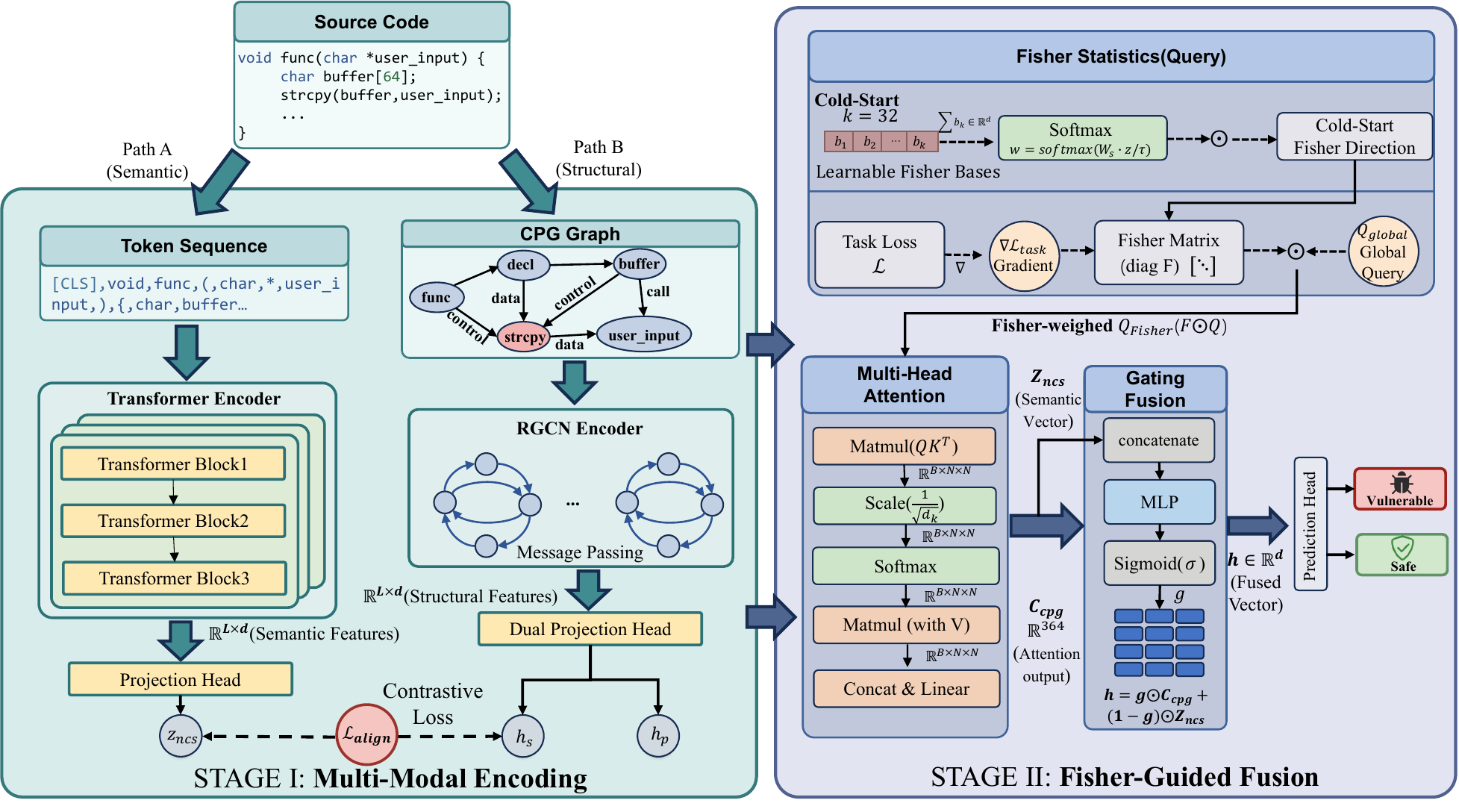}
\vspace{-0.2cm}
\caption{Overall architecture of TaCCS-DFA}
\label{fig:framework}
\vspace{-0.2cm}
\end{figure*}

This chapter elaborates on the construction details of the TaCCS-DFA framework. To address the issues of feature redundancy and modal asymmetry in code multimodal analysis, this framework establishes a task-conditioned complementary fusion strategy based on information geometry. The overall architecture is shown in Figure~\ref{fig:framework}, with its core being the use of the online approximated FIM as a prior to dynamically guide the attention mechanism in retrieving key structural subspaces in the auxiliary modality (CPG) that are complementary to the dominant modality (NCS).
\subsection{Problem Formulation}
Software vulnerability detection can be formalized as a binary classification problem based on multimodal data. Given a dataset $\mathcal{D} = \{(c_i, y_i)\}_{i=1}^N$, where $c_i$ denotes a source code function and $y_i \in \{0, 1\}$ is the corresponding vulnerability label (1 indicates the presence of a vulnerability). Each sample $c_i$ can be modeled as two heterogeneous views: the NCS view $x_{\text{ncs}}$ and the CPG view $\mathcal{G}_{\text{cpg}}$.

The model’s overall mapping function $\mathcal{F}_{\Theta}: (x_{\text{ncs}}, \mathcal{G}_{\text{cpg}}) \mapsto \hat{y}$ aims to learn parameters $\Theta$ such that the cross-entropy loss $\mathcal{L}_{\text{ce}}$ between the predicted probability distribution $p_{\Theta}(y|c_i)$ and the true distribution is minimized. In light of the modal asymmetry phenomenon revealed by the previous analysis, this paper adopts a fusion strategy with NCS as the dominant modality and CPG as the auxiliary modality. To precisely extract effective information from the auxiliary modality, we utilize the Fisher information matrix to identify task-sensitive subspaces, thereby screening out key features complementary to NCS from CPG.
\subsection{Unimodal Feature Encoding}
For the NCS view, pretrained models such as CodeBERT or CodeT5 are employed for encoding, and the final layer hidden states are taken as the semantic representation $\mathbf{H}_{\text{ncs}} \in \mathbb{R}^{L \times d}$, with the [CLS] token vector $\mathbf{h}_{\text{ncs}}^{\text{cls}} \in \mathbb{R}^{d}$ serving as the global representation.

For the CPG view $\mathcal{G}_{\text{cpg}} = (\mathcal{V}, \mathcal{E})$, the RGCN\cite{schlichtkrull2018modeling} is used to process the heterogeneous graph structure with multiple edge types. RGCN is an extension of GCN that can handle heterogeneous graphs containing multiple edge types by learning independent weight matrices for different relation types to aggregate neighbor information. The node update rule is:
\begin{equation}
\mathbf{h}_i^{(l+1)} = \sigma\left( \sum_{r \in \mathcal{R}} \sum_{j \in \mathcal{N}_r(i)} \frac{1}{c_{i,r}} \mathbf{W}_r^{(l)} \mathbf{h}_j^{(l)} + \mathbf{W}_0^{(l)} \mathbf{h}_i^{(l)} \right)
\end{equation}
After $K$ layers of aggregation, the node representation matrix $\mathbf{H}_{\text{cpg}} \in \mathbb{R}^{|\mathcal{V}| \times d}$ is obtained.
\subsection{Stage I: Cross-Modal Alignment}
There exists a significant modal gap between the feature distributions generated by pretrained language models and graph neural networks. To establish semantic correspondence between the two, we map the heterogeneous representations to a shared metric space through contrastive learning.

\textbf{Dual Projection Heads.}
We design independent nonlinear projection heads for the modalities to map the features into low-dimensional contrastive space:
\begin{equation}
\mathbf{z}_{\text{ncs}} = \text{MLP}_{\text{ncs}}(\mathbf{h}_{\text{ncs}}^{\text{cls}}), \quad \mathbf{z}_{\text{cpg}} = \text{MLP}_{\text{cpg}}(\mathbf{h}_{\text{cpg}}^{\text{pool}})
\end{equation}
where the projection vectors $\mathbf{z} \in \mathbb{R}^{d'}$ are $L_2$-normalized and lie on the unit hypersphere.

\textbf{Contrastive Alignment.}
The InfoNCE loss\cite{oord2018representation} is adopted to constrain the consistency of cross-modal representations:
\begin{equation}
\mathcal{L}_{\text{align}} = -\log \frac{\exp(\mathbf{z}_{\text{ncs}}^{\top} \mathbf{z}_{\text{cpg}} / \tau)}{\sum_{k=1}^{B} \exp(\mathbf{z}_{\text{ncs}}^{\top} \mathbf{z}_{\text{cpg}}^{(k)} / \tau)}
\end{equation}
where $\tau$ is the temperature coefficient. To increase the diversity of negative samples, a cross-batch memory queue (XBM) with capacity $Q$ is introduced. This stage lays the geometric foundation for subsequent Fisher-guided cross-modal interaction.

\subsection{Stage II: Dynamic Fisher Attention for Task-Conditioned Complementary Fusion}
The standard cross-attention mechanism $\text{Attn}(\mathbf{Q}, \mathbf{K}, \mathbf{V})$ typically computes weights based on content-based similarity. This mechanism lacks awareness of the task objective and cannot distinguish task-relevant effective features from redundant noise in the auxiliary modality, potentially introducing noise into the fusion results. To address this issue, this paper proposes Dynamic Fisher Attention (DFA). The core idea of this module is that the generation of the attention query vector (Query) should not depend solely on the input features themselves but should be driven by the sensitivity of the features to the task objective.

\textbf{Incremental Fisher Estimation.}
Geometrically, the FIM characterizes the sensitivity of the model's predicted distribution to small perturbations in the feature space. For a feature representation $\mathbf{h}$, its FIM is defined as
\begin{equation}
\mathbf{F} = \mathbb{E}\big[\nabla_{\mathbf{h}} \log p(y|\mathbf{h}) \nabla_{\mathbf{h}} \log p(y|\mathbf{h})^\top\big] \, .
\end{equation}
Feature directions with high Fisher information correspond to the subspaces where the loss function has the largest curvature, i.e., variations in features along these directions have a decisive influence on the final classification decision.

However, explicitly constructing and frequently updating the full $d\times d$ FIM in high-dimensional deep networks faces severe computational bottlenecks. Accurately computing second-order statistics in each training iteration requires $O(d^2)$ operations, and performing full-matrix eigendecomposition to extract the Fisher principal subspace has $O(d^3)$ time complexity. Such high computational costs make real-time updating of Fisher information within the training loop of deep neural networks unscalable.To address this challenge, an efficient method is needed to track the principal eigenspace of the FIM. This paper introduces Oja's Rule~\cite{oja1982simplified}, an online principal component analysis algorithm based on Hebbian learning~\cite{hebb1949organization}, which can incrementally approximate the top $k$ principal eigenvectors of the FIM during training. This method avoids full-matrix eigendecomposition through iterative updates, reducing the space complexity to $O(dk)$ and the per-step update complexity to $O(dk)$ as well.

Let $\mathbf{U}_t \in \mathbb{R}^{d \times k}$ be the estimated orthogonal basis of the Fisher subspace at the $t$-th iteration. In each training batch, the gradient of the cross-entropy loss with respect to the NCS features $\mathbf{G}_t = \nabla_{\mathbf{H}_{\text{ncs}}} \mathcal{L}_{\text{ce}} \in \mathbb{R}^{L \times d}$ is computed and compressed into $\mathbf{g}_t = \text{Pool}(\mathbf{G}_t) \in \mathbb{R}^{d}$ via mean pooling. The Oja update rules are as follows:
\begin{equation}
\mathbf{y}_t = \mathbf{U}_t^\top \mathbf{g}_t
\end{equation}
\begin{equation}
\mathbf{U}_{t+1} = \mathbf{U}_t + \eta_t \big(\mathbf{g}_t \mathbf{y}_t^\top - \mathbf{U}_t \mathbf{y}_t \mathbf{y}_t^\top\big)
\end{equation}
where $\eta_t$ is the learning rate. It is worth noting that for the cross-entropy loss $\mathcal{L}_{\text{ce}}=-\log p_{\theta}(y|x)$, we have $\nabla_{\mathbf{h}}\log p_{\theta}(y|x) = -\nabla_{\mathbf{h}}\mathcal{L}_{\text{ce}}$, so approximating the Fisher second-moment with $\mathbf{g}_t\mathbf{g}_t^\top$ is consistent in the outer-product sense. This update process tracks the top $k$ principal eigenvectors of the FIM without explicitly constructing the full Fisher matrix $\mathbf{F}$. To maintain column orthogonality of $\mathbf{U}_t$, we perform orthogonalization on $\mathbf{U}_{t+1}$ after each update.

\textbf{Task-Conditioned Query Generation.}
After obtaining the Fisher subspace $\mathbf{U}$, we inject it as prior knowledge into the generation process of query vectors. The conventional query vector $\mathbf{Q} = \mathbf{H}_{\text{ncs}}\mathbf{W}_q$ encodes only semantic information. In DFA, we explicitly enhance the components lying in high Fisher information feature directions:
\begin{equation}
\mathbf{Q}_{\text{dfa}} = \left( \mathbf{H}_{\text{ncs}} + \text{LayerNorm}(\mathbf{H}_{\text{ncs}} \mathbf{U} \mathbf{U}^\top) \right)\mathbf{W}_q
\end{equation}
where $\mathbf{U} \mathbf{U}^\top$ is the projection matrix onto the Fisher principal subspace. By adding the original features to their projections along the Fisher-sensitive directions, the constructed $\mathbf{Q}_{\text{dfa}}$ acquires task-aware properties: it tends to search in the auxiliary modality for structural clues that can explain the high-sensitivity semantic features, rather than merely matching semantically similar nodes.

\textbf{Complementary Subspace Attention.}
Using the task-aware query vector $\mathbf{Q}_{\text{dfa}}$, we perform multi-head attention on the CPG representations to extract complementary structural features. Since the cross-modal alignment training in Stage I aligns the two modalities in the same $d$-dimensional semantic coordinate system, the Fisher principal subspace computed from NCS features can be directly applied to filter CPG representations. To ensure that cross-modal interaction occurs only within the task-sensitive subspace, we apply subspace filtering to the auxiliary modality features. Let $\mathbf{P}=\mathbf{U}\mathbf{U}^\top \in \mathbb{R}^{d\times d}$ be the orthogonal projection matrix onto the Fisher principal subspace $\mathcal{S}_{\text{fisher}}=\mathrm{span}(\mathbf{U})$. First, the CPG node representations are projected as:
\begin{equation}
\mathbf{H}_{\text{cpg}}^{\parallel} = \mathbf{H}_{\text{cpg}} \mathbf{P} = \mathbf{H}_{\text{cpg}}\mathbf{U}\mathbf{U}^\top
\end{equation}
Subsequently, the key and value vectors are constructed solely based on $\mathbf{H}_{\text{cpg}}^{\parallel}$:
\begin{equation}
\mathbf{H}_{\text{comp}} = \text{Softmax}\left( \frac{\mathbf{Q}_{\text{dfa}} (\mathbf{H}_{\text{cpg}}^{\parallel}\mathbf{W}_k)^\top}{\sqrt{d_k}} \right) \mathbf{H}_{\text{cpg}}^{\parallel}\mathbf{W}_v
\end{equation}
Under this mechanism, the distribution of attention weights is no longer based on static semantic alignment but reflects the measure of task criticality. Since $\mathbf{P}$ eliminates components falling in orthogonal complement space $\mathcal{S}_{\perp}$, a large amount of topology noise in CPG that is insensitive to the task objective is filtered before entering the attention, thereby simultaneously suppressing its influence on logits and value vectors. The final output $\mathbf{H}_{\text{comp}}$ retains only structural features in CPG that are highly relevant to the current discriminative task.

\subsection{Adaptive Gating Fusion}
Software vulnerabilities exhibit significant heterogeneity; different samples vary in degree of dependence on structural information enhancement. Simple buffer overflows may be identifiable solely through lexical patterns, whereas complex UAF vulnerabilities highly depend on data flow graphs. Forcibly fusing graph features may introduce unnecessary interference to simple samples. To address this, we design a lightweight adaptive gating unit that dynamically regulates the modal fusion ratio according to each sample’s own semantic complexity.

Using the global semantic vector of NCS $\mathbf{h}_{\text{ncs}}^{\text{cls}}$ and the pooled vector of complementary structural features $\mathbf{h}_{\text{comp}}^{\text{pool}}$, the gating coefficient $\alpha \in [0, 1]$ is computed as:
\begin{equation}
\alpha = \sigma\left( \mathbf{w}_g^\top [\mathbf{h}_{\text{ncs}}^{\text{cls}} \parallel \mathbf{h}_{\text{comp}}^{\text{pool}}] + b_g \right)
\end{equation}
The final multimodal representation $\mathbf{h}_{\text{final}}$ is obtained via residual connection:
\begin{equation}
\mathbf{h}_{\text{final}} = \mathbf{h}_{\text{ncs}}^{\text{cls}} + \alpha \cdot \mathbf{W}_o \mathbf{h}_{\text{comp}}^{\text{pool}}
\end{equation}
This gating mechanism acts as a learnable control valve: when the confidence provided by NCS is sufficient to perform discrimination, the model can automatically decrease the value of $\alpha$ to shield graph-modal noise; conversely, when confidence is insufficient, the model increases $\alpha$ to introduce structured evidence.

\textbf{Training Objectives.}
TaCCS-DFA adopts an end-to-end joint training strategy. The total loss function $\mathcal{L}_{\text{total}}$ consists of the cross-entropy loss $\mathcal{L}_{\text{ce}}$ of the main task and the auxiliary cross-modal alignment loss $\mathcal{L}_{\text{align}}$ with a weighting factor:
\begin{equation}
\mathcal{L}_{\text{total}} = \mathcal{L}_{\text{ce}}(\hat{y}, y) + \beta \cdot \mathcal{L}_{\text{align}}
\end{equation}
where $\beta$ is the balancing coefficient that controls the strength of the alignment constraint. In the early training phase, $\mathcal{L}_{\text{align}}$ dominates to rapidly align the feature spaces; as training progresses, the model gradually focuses on $\mathcal{L}_{\text{ce}}$ to optimize the decision boundary while utilizing Fisher information for refined feature enhancement.

In implementation, this paper employs a two-stage scheduling strategy that sets different alignment weights $\beta$ across training stages: Stage~I uses a larger weight to strengthen cross-modal alignment, while Stage~II reduces the weight to optimize the decision boundary while retaining the alignment constraint.

\subsection{Theoretical Robustness Analysis}
To verify the theoretical advantages of the TaCCS-DFA method for suppressing modal noise, we analyze the risk upper bound under input perturbations. We prove that, compared to the full-spectrum standard attention mechanism, DFA can significantly reduce the impact of auxiliary modality noise on the final prediction by constraining attention within the Fisher principal subspace.

\textbf{Noise Model and Decomposition.}
Assume that the auxiliary modality CPG feature $\mathbf{H}_{\text{cpg}}$ is contaminated by additive noise $\bm{\Delta}$, i.e., $\tilde{\mathbf{H}}_{\text{cpg}} = \mathbf{H}_{\text{cpg}} + \bm{\Delta}$, and $\|\bm{\Delta}\|_F \le \varepsilon$.Using the orthogonal basis $\mathbf{U} \in \mathbb{R}^{d \times k}$ of the Fisher principal subspace estimated by Oja's Rule, we decompose the noise into a parallel component $\bm{\Delta}_\parallel$ (lying in the sensitive subspace $\mathcal{S}_{\text{fisher}}$) and a perpendicular component $\bm{\Delta}_\perp$ (lying in the non-sensitive subspace $\mathcal{S}_{\perp}$):
\begin{equation}
\bm{\Delta} = \bm{\Delta}_\parallel + \bm{\Delta}_\perp, \quad \text{where } \bm{\Delta}_\parallel = \bm{\Delta}\mathbf{U}\mathbf{U}^\top.
\end{equation}

\textbf{Robustness Bound.}
In the robustness analysis of this paper, the \textbf{Risk Upper Bound} is defined as the provable upper limit of the model output deviation in the worst case when the input features are subject to bounded perturbations\cite{hein2017formal}. Formally, given a perturbation $\bm{\Delta}$ satisfying $\|\bm{\Delta}\|_F \le \varepsilon$, the risk upper bound of the mapping $\mathcal{F}$ is a constant $C(\varepsilon)$ such that $\|\mathcal{F}(\tilde{\mathbf{H}}) - \mathcal{F}(\mathbf{H})\|_F \le C(\varepsilon)$\cite{tsuzuku2018lipschitz}. This metric characterizes the sensitivity of the attention fusion mechanism to input noise: the smaller the upper bound, the stronger the model's robustness to perturbations\cite{gouk2021lipschitz}.

In standard cross-attention, the query vector $\mathbf{Q}$ may align with noise in any direction, causing the output error upper bound to depend on the full noise $\|\bm{\Delta}\|_F$.In TaCCS-DFA, the query vector $\mathbf{Q}_{\text{dfa}}$ is generated under the guidance of the Fisher principal subspace, and its column space lies approximately within $\mathcal{S}_{\text{fisher}}$.Based on the rapid spectral decay property of the Fisher spectrum, the high-sensitivity dimension $k$ is much smaller than the total feature dimension $d$.Under the Lipschitz continuity assumption, we derive the following theorem:
\begin{theorem}[Compactness of the DFA Perturbation Upper Bound]
\label{thm:bound}
Let the Lipschitz constant of the attention mechanism be $L$. For any input perturbation satisfying $\|\bm{\Delta}\|_F \le \varepsilon$, the output error upper bound of the full-spectrum attention mechanism $\mathcal{F}_{\text{full}}$ is:
\begin{equation}
\|\mathcal{F}_{\text{full}}(\tilde{\mathbf{H}}_{\text{cpg}}) - \mathcal{F}_{\text{full}}(\mathbf{H}_{\text{cpg}})\|_F \le L \cdot \varepsilon
\end{equation}
For the TaCCS-DFA mechanism $\mathcal{F}_{\text{dfa}}$, the effective input perturbation is determined solely by the Fisher principal subspace component $\bm{\Delta}_{\parallel}$, yielding the deterministic upper bound:
\begin{equation}
\|\mathcal{F}_{\text{dfa}}(\tilde{\mathbf{H}}_{\text{cpg}}) - \mathcal{F}_{\text{dfa}}(\mathbf{H}_{\text{cpg}})\|_F \le L \cdot \|\bm{\Delta}_{\parallel}\|_F \le L \cdot \varepsilon.
\end{equation}
When the isotropic noise assumption is further adopted, the expected error upper bound satisfies:
\begin{equation}
\mathbb{E}\left[\|\mathcal{F}_{\text{dfa}}(\tilde{\mathbf{H}}_{\text{cpg}}) - \mathcal{F}_{\text{dfa}}(\mathbf{H}_{\text{cpg}})\|_F\right] \le L \cdot \sqrt{\frac{k}{d}} \cdot \varepsilon
\end{equation}
where $\sqrt{k/d}$ is the noise suppression factor. Since $k \ll d$, DFA significantly tightens the error upper bound.
\end{theorem}

\textbf{Geometric Interpretation.}
Theorem \ref{thm:bound} reveals the geometric essence of the Fisher-guided mechanism: it acts as a task-conditioned low-pass filter for the feature manifold.In Complementary Subspace Attention, we explicitly project the auxiliary modality representations onto the Fisher principal subspace: $\mathbf{H}_{\text{cpg}}^{\parallel}=\mathbf{H}_{\text{cpg}}\mathbf{U}\mathbf{U}^\top$. Denote $\mathcal{S}_{\text{fisher}} = \mathrm{span}(\mathbf{U})$ as the Fisher principal subspace spanned by the column vectors of $\mathbf{U}$, and $\mathcal{S}_{\perp}$ as its orthogonal complement. Due to their orthogonality, the noise component $\bm{\Delta}_\perp$ satisfies $\bm{\Delta}_\perp \mathbf{U}\mathbf{U}^\top=\mathbf{0}$, so that the attention logits and output of DFA are affected only by $\bm{\Delta}_\parallel$. This means DFA automatically filters out noise perturbations that are insensitive to the task objective (low Fisher information), allowing only a minimal amount of noise to propagate through the sensitive subspace.In contrast, the standard attention mechanism cannot distinguish effective structural information from topological noise, causing noise to propagate across the entire feature space. The detailed proof is provided in Appendix \ref{appendix:proof}.

\section{Experiments}
This section aims to comprehensively evaluate the effectiveness, robustness, and computational efficiency of TaCCS-DFA in the software vulnerability detection task through multidimensional empirical studies. Specifically, our experimental design is dedicated to answering the following four core research questions:
\begin{itemize}
\item \textbf{RQ1 (Performance Advantage)}: Compared with existing unimodal and multimodal fusion baselines, can TaCCS-DFA significantly improve detection performance while maintaining a low false positive rate, particularly in scenarios with extremely imbalanced classes?
\item \textbf{RQ2 (Mechanism Validity)}: Is the Fisher information-guided mechanism the key source of performance improvement? Does its effectiveness stem from the introduction of geometric information rather than an increase in the number of parameters? Is the true structure of CPG indispensable?
\item \textbf{RQ3 (Explainability)}: Can Fisher attention precisely focus on vulnerability causal paths, and is its subspace filtering behavior consistent with the theoretical analysis?
\item \textbf{RQ4 (Efficiency and Overhead)}: Can the training overhead, inference latency, and memory footprint of TaCCS-DFA meet the requirements of practical deployment?
\end{itemize}


\subsection{Experimental Setup}
\textbf{Datasets and Metrics.}
We evaluate on four benchmark datasets, all derived from real-world software projects. BigVul~\cite{fan2020ac} is constructed based on public CVE fix commits, with vulnerable samples accounting for 5.0\%, exhibiting severe class imbalance; PrimeVul~\cite{ding2025vulnerability} is a recently proposed high-quality dataset that has undergone rigorous data deduplication and label verification, with a vulnerability ratio of only 3.0\% and covering more than 140 CWE types, enabling a more realistic simulation of actual code auditing scenarios and representing the most challenging benchmark among the four datasets; Devign~\cite{zhou2019devign} is likewise sourced from real security fix commits, with a relatively balanced positive-to-negative sample distribution, making it suitable for assessing method effectiveness under approximately balanced distributions; ReVeal~\cite{chakraborty2020deep} originates from real vulnerability fix records across multiple open-source projects, with paired vulnerable and patched code and reliable labels. These four datasets span a range of data distributions from approximately balanced to highly imbalanced; detailed statistics for each dataset are provided in Table~\ref{tab:dataset_stats}. Evaluation metrics include Precision, Recall, Accuracy, and F1-score. In addition, to measure the reliability of model prediction confidence, we introduce the Expected Calibration Error (ECE)~\cite{guo2017calibration} metric. All experiments adopt the official splits provided by the original papers of each dataset, and samples that fail Joern parsing are filtered during preprocessing; the above processing is applied identically to the proposed method and all baseline methods.

\begin{table}[t]
  \centering
  \caption{Dataset statistics}
  \vspace{-0.3cm}
  \label{tab:dataset_stats}
  \scriptsize
  \setlength{\tabcolsep}{2pt}
  \begin{tabular}{@{}lcrrrrrc@{}}
  \toprule
  \textbf{Dataset} & \textbf{Language} & \textbf{Train} & \textbf{Val} & \textbf{Test} & \textbf{Total} & \textbf{\#Vul} & \textbf{Vul Ratio} \\
  \midrule
  BigVul & C/C++ & 150,908 & 33,049 & 33,050 & 217,007 &10,895 & 5.0\% \\
  PrimeVul & C/C++ & 184,427 & 25,430 & 25,911 & 235,768 & 6,968 & 3.0\% \\
  Devign & C & 21,854 & 2,732 & 2,732 & 27,318 &12,460 & 45.6\% \\
  ReVeal & C & 18,187 & 2,273 & 2,274 & 22,734 &2,240 & 9.9\% \\
  \bottomrule
  \end{tabular}
\vspace{-0.2cm}
\end{table}

\textbf{Vulnerability Type Coverage.}
The detailed CWE label coverage and vulnerability type distribution for BigVul and PrimeVul are provided in Appendix~\ref{appendix:dataset_cwe}. In brief, both datasets are dominated by memory-safety vulnerabilities, which matches the target scenarios emphasized in our evaluation.

\textbf{Baselines.}
We compare \mymethod with three categories of baseline methods. The first category consists of unimodal methods, including the text-based CodeBERT and CodeT5, as well as the graph-based RGCN. The second category comprises basic fusion strategies, covering feature concatenation, cross-attention, and gated fusion. To exclude the impact of parameter scale on performance, we specifically design a Concat+MLP variant whose parameter count is matched to that of \mymethod. The third category includes prior state-of-the-art methods such as Devign, GraphCodeBERT, and the Vul-LMGNNs series.

\textbf{Implementation Details.}
All experiments are conducted on four NVIDIA 3090 GPUs using AdamW for 15 epochs. The complete training configuration and hyperparameters are provided in Appendix~\ref{appendix:implementation_details}.





\subsection{RQ1: Effectiveness and Asymmetry Mitigation}
Tables~\ref{tab:imbalanced_results} and~\ref{tab:balanced_results} present the end-to-end detection performance on the four benchmark datasets. The experimental results indicate that TaCCS-DFA achieves significant improvements across different backbone networks. When CodeT5-Base is used as the backbone, the model attains an F1 score of 0.8780 on BigVul. Under the same CodeT5-Base backbone, compared with Vul-LMGNNs, TaCCS-DFA improves precision from 87.88\% to 92.31\%, recall from 67.44\% to 83.72\%, and F1 from 76.32\% to 87.80\%. In the BigVul test set containing approximately 1,652 vulnerable samples, the recall improvement alone implies that approximately 269 additional vulnerable functions are correctly identified, demonstrating simultaneous mitigation of false negatives and false positives. This trend consistently appears on the PrimeVul, Devign, and ReVeal datasets, strongly confirming the superiority of the proposed task-oriented fusion paradigm.

Existing multimodal baselines often exhibit severe precision-recall imbalance when processing imbalanced data. For example, GraphCodeBERT achieves a high precision (0.9655) but a relatively low recall (0.6512), indicating a substantial number of false negatives; TaCCS-DFA, by filtering CPG noise via Fisher information, significantly improves recall while maintaining high precision, thereby effectively reducing the threat of false negatives to security auditing.

In terms of generalization, TaCCS-DFA achieves F1 scores of 0.6460 and 0.5381 on Devign and ReVeal, respectively. On the PrimeVul dataset, which has the strictest labels and a vulnerability ratio of only 3.0\%, TaCCS-DFA likewise attains the best F1 score of 0.2525 among all methods, further validating the robustness of the method under highly imbalanced distributions. Few-shot LLM baselines remain clearly below task-specific multimodal models and exhibit a high-recall, low-precision imbalance; detailed results are provided in Appendix~\ref{appendix:llm_results} and the prompt template is given in Appendix~\ref{appendix:llm-fewshot-prompt}.

Appendix~\ref{appendix:main_results_profile} provides a unified metrics-profile visualization of Precision, Recall, Accuracy, and F1 across the four benchmark datasets. Consistent with Tables~\ref{tab:imbalanced_results} and~\ref{tab:balanced_results}, TaCCS-DFA maintains a relatively balanced precision-recall trade-off across different data distributions and achieves the strongest overall F1 performance, indicating good robustness and generalization capability.

\begin{table*}[htbp]
\centering
\scriptsize
\setlength{\tabcolsep}{2pt}
\caption{Comparison of Vulnerability Detection Performance of Various Models on the Two Highly Class-Imbalanced Datasets BigVul and PrimeVul}
\vspace{-0.3cm}
\label{tab:imbalanced_results}
\sisetup{
  table-format=2.2,
  round-mode=places,
  round-precision=2,
  detect-weight,
  detect-display-math=true
}
\resizebox{\linewidth}{!}{%
\begin{tabular}{@{}l l *{4}{S[table-format=2.2]} *{4}{S[table-format=2.2]}@{}}
\toprule
& & \multicolumn{4}{c}{\textbf{BigVul Dataset}} & \multicolumn{4}{c}{\textbf{PrimeVul Dataset}} \\
\cmidrule(lr){3-6} \cmidrule(lr){7-10}
& \textbf{Model} & {P (\%)} & {R (\%)} & {ACC (\%)} & {F1 (\%)} & {P (\%)} & {R (\%)} & {ACC (\%)} & {F1 (\%)} \\
\midrule
\multicolumn{10}{l}{\cellcolor[HTML]{9ECFD4}{\textit{Unimodal Baselines}}} \\
\Xhline{0.5pt}
& NCS (CodeBERT) & 72.92 & 56.45 & 96.01 & 63.64 & 14.57 & 36.43 & 93.86 & 20.81 \\
& CPG (RGCN) & 12.31 & 37.21 & 73.30 & 18.50 & 13.16 & 37.16 & 93.18 & 19.44 \\
& CodeT5-Small & 60.34 & 81.40 & 94.13 & 69.31 & 14.86 & 34.24 & 94.20 & 20.73 \\
& CodeT5-Base & 71.11 & 74.42 & 95.45 & 72.73 & 25.51 & 20.58 & 96.91 & 22.78 \\
& CodeLLaMA-7B (fine-tuned) & 74.71 & 82.36 & 98.59 & 78.35 & 9.57 & 37.70 & 90.62 & 15.26 \\
\Xhline{0.5pt}
\multicolumn{10}{l}{\cellcolor[HTML]{9ECFD4}{\textit{Fusion Baselines$^\dagger$}}} \\
\Xhline{0.5pt}
& ConcatFusion & 67.44 & 68.24 & 94.89 & 69.44 & 25.05 & 21.49 & 96.84 & 23.14 \\
& Concat+MLP & 92.86 & 60.47 & 96.40 & 73.24 & 21.50 & 24.59 & 96.34 & 22.94 \\
& Cross-Attention & 89.66 & 60.47 & 96.21 & 72.22 & 17.75 & 26.96 & 95.61 & 21.40 \\
& Gated Fusion & 78.95 & 69.77 & 96.02 & 74.07 & 25.99 & 15.48 & 97.15 & 19.41 \\
\Xhline{0.5pt}
\multicolumn{10}{l}{\cellcolor[HTML]{9ECFD4}{\textit{Representative Baselines}}} \\
\Xhline{0.5pt}
& Devign & 18.03 & 25.58 & 84.47 & 21.15 & 9.80 & 32.06 & 91.96 & 15.01 \\
& VulMPFF & 25.00 & 18.60 & 88.83 & 21.33 & 23.00 & 13.11 & 97.10 & 16.70 \\
& VulBERTa-MLP & 19.44 & 32.56 & 83.52 & 24.35 & 21.32 & 17.12 & 96.76 & 18.99 \\
& VulBERTa-CNN & 17.91 & 55.81 & 75.57 & 27.12 & 20.62 & 24.04 & 96.27 & 22.20 \\
& GraphCodeBERT & 96.55 & 65.12 & 96.97 & 77.78 & 17.05 & 29.87 & 95.23 & 21.71 \\
& Vul-LMGNNs (GraphCodeBERT) & 90.62 & 67.44 & 96.78 & 77.33 & 12.27 & 25.32 & 94.34 & 16.53 \\
& Vul-LMGNNs (CodeBERT) & 82.86 & 67.44 & 96.21 & 74.36 & 29.19 & 15.85 & 97.28 & 20.54 \\
& Vul-LMGNNs (CodeT5-Small) & 86.84 & 76.74 & 97.16 & 81.48 & 27.65 & 18.03 & 97.14 & 21.83 \\
& Vul-LMGNNs (CodeT5-Base) & 87.88 & 67.44 & 96.59 & 76.32 & 35.71 & 11.84 & 97.58 & 17.78 \\
\Xhline{0.5pt}
\multicolumn{10}{l}{\cellcolor[HTML]{9ECFD4}{\textit{Ours}}} \\
\Xhline{0.5pt}
& \textbf{\mymethod (CodeBERT)} & \textbf{96.67} & \textbf{67.44} & \textbf{97.16} & \textbf{79.45} & \textbf{22.07} & \textbf{29.51} & \textbf{96.13} & \textbf{25.25} \\
& \textbf{\mymethod (CodeT5-Small)} & \textbf{94.44} & \textbf{79.07} & \textbf{97.92} & \textbf{86.08} & \textbf{17.93} & \textbf{29.33} & \textbf{95.46} & \textbf{22.25} \\
& \textbf{\mymethod (CodeT5-Base)} & \textbf{92.31} & \textbf{83.72} & \textbf{98.11} & \textbf{87.80} & \textbf{23.50} & \textbf{23.86} & \textbf{96.60} & \textbf{23.68} \\
\bottomrule
\end{tabular}%
}
\vspace{-0.2cm}
\flushleft{\small
$^\dagger$ All Fusion Baselines use CodeBERT as the backbone.
}
\vspace{-0.2cm}
\end{table*}

\begin{table*}[htbp]
\centering
\scriptsize
\setlength{\tabcolsep}{2pt}
\caption{Comparison of Vulnerability Detection Performance of Various Models on the Devign and ReVeal Datasets}
\vspace{-0.3cm}
\label{tab:balanced_results}
\sisetup{
  table-format=2.2,
  round-mode=places,
  round-precision=2,
  detect-weight,
  detect-display-math=true
}
\resizebox{\linewidth}{!}{%
\begin{tabular}{@{}l l *{4}{S[table-format=2.2]} *{4}{S[table-format=2.2]}@{}}
\toprule
& & \multicolumn{4}{c}{\textbf{Devign Dataset}} & \multicolumn{4}{c}{\textbf{ReVeal Dataset}} \\
\cmidrule(lr){3-6} \cmidrule(lr){7-10}
& \textbf{Model} & {P (\%)} & {R (\%)} & {ACC (\%)} & {F1 (\%)} & {P (\%)} & {R (\%)} & {ACC (\%)} & {F1 (\%)} \\
\midrule
\multicolumn{10}{l}{\cellcolor[HTML]{9ECFD4}{\textit{Unimodal Baselines}}} \\
\Xhline{0.5pt}
& NCS (CodeBERT) & 61.03 & 58.10 & 63.93 & 59.53 & 44.44 & 47.93 & 88.97 & 46.12 \\
& CPG (RGCN) & 54.98 & 42.63 & 57.56 & 48.02 & 25.97 & 59.70 & 78.23 & 36.20 \\
& CodeT5-Small & 64.35 & 55.39 & 65.62 & 59.53 & 47.81 & 40.73 & 90.50 & 43.99 \\
& CodeT5-Base & 64.27 & 57.12 & 65.92 & 60.48 & 50.69 & 39.76 & 90.94 & 44.56 \\
& CodeLLaMA-7B (fine-tuned) & 67.96 & 49.00 & 65.96 & 56.94 & 33.05 & 50.87 & 84.61 & 40.07 \\
\Xhline{0.5pt}

\multicolumn{10}{l}{\cellcolor[HTML]{9ECFD4}{\textit{Fusion Baselines$^\dagger$}}} \\
\Xhline{0.5pt}
& ConcatFusion & 72.21 & 38.84 & 66.84 & 50.50 & 43.69 & 47.09 & 88.38 & 45.33 \\
& Concat+MLP & 68.32 & 44.47 & 64.99 & 53.87 & 48.08 & 49.75 & 89.24 & 48.89 \\
& Cross-Attention & 64.96 & 45.52 & 63.65 & 53.53 & 79.76 & 33.33 & 92.23 & 47.02 \\
& Gated Fusion & 63.45 & 53.60 & 64.46 & 58.11 & 55.06 & 42.23 & 90.56 & 47.80 \\
\Xhline{0.5pt}

\multicolumn{10}{l}{\cellcolor[HTML]{9ECFD4}{\textit{Representative Baselines}}} \\
\Xhline{0.5pt}
& Devign & 56.96 & 56.25 & 57.66 & 56.60 & 36.65 & 31.55 & 87.49 & 33.91 \\
& VulMPFF & 54.49 & 71.32 & 59.42 & 61.78 & 25.34 & 82.59 & 73.03 & 38.79 \\
& VulBERTa-MLP & 62.71 & 56.22 & 64.75 & 59.29 & 36.79 & 35.90 & 88.48 & 36.34 \\
& VulBERTa-CNN & 63.11 & 53.12 & 64.42 & 57.29 & 34.46 & 38.76 & 87.64 & 36.48 \\
& GraphCodeBERT & 64.37 & 54.38 & 64.80 & 58.96 & 41.67 & 41.81 & 89.25 & 41.74 \\
& Vul-LMGNNs (GraphCodeBERT) & 64.73 & 57.77 & 66.33 & 61.01 & 55.12 & 43.41 & 91.58 & 48.57 \\
& Vul-LMGNNs (CodeBERT) & 64.53 & 56.34 & 65.70 & 60.16 & 57.09 & 46.45 & 90.80 & 51.22 \\
& Vul-LMGNNs (CodeT5-Small) & 63.45 & 64.20 & 66.77 & 63.82 & 50.76 & 50.41 & 90.98 & 50.58 \\
& Vul-LMGNNs (CodeT5-Base) & 64.73 & 62.20 & 67.27 & 63.44 & 54.89 & 51.41 & 91.68 & 53.09 \\
\Xhline{0.5pt}

\multicolumn{10}{l}{\cellcolor[HTML]{9ECFD4}{\textit{Ours}}} \\
\Xhline{0.5pt}
& \textbf{\mymethod (CodeBERT)} & \textbf{59.54} & \textbf{65.44} & \textbf{64.30} & \textbf{62.35} & \textbf{51.43} & \textbf{53.73} & \textbf{89.96} & \textbf{52.55} \\
& \textbf{\mymethod (CodeT5-Small)} & \textbf{57.03} & \textbf{70.44} & \textbf{62.00} & \textbf{63.03} & \textbf{39.34} & \textbf{70.65} & \textbf{85.69} & \textbf{50.53} \\
& \textbf{\mymethod (CodeT5-Base)} & \textbf{57.40} & \textbf{73.86} & \textbf{62.77} & \textbf{64.60} & \textbf{51.60} & \textbf{56.22} & \textbf{90.02} & \textbf{53.81} \\
\bottomrule
\end{tabular}%
}
\vspace{-0.2cm}
\flushleft{\small
$^\dagger$ All Fusion Baselines use CodeBERT as the backbone.
}
\vspace{-0.2cm}
\end{table*}

\subsection{RQ2: Validity of Fisher Guidance}
To gain a deeper understanding of the intrinsic mechanisms driving the performance improvement of TaCCS-DFA, we design systematic ablation studies on the BigVul dataset, as in Table~\ref{tab:ablation}. The experiments revolve around three core questions: whether Fisher guidance is the key factor, whether its effectiveness stems from the introduced geometric information rather than an increase in parameter count, and whether the true structure of CPG is indispensable.

\textbf{Core Component Ablation.}
Removing Fisher guidance causes the F1 score to drop from 79.45\% to 77.78\%, while removing adaptive gating reduces F1 to 75.68\%. Replacing the Fisher projection matrix with a random orthogonal matrix $B_{\text{rand}}$ causes a larger decline to 73.68\%, showing that the gain comes from task-relevant geometric directions rather than the projection operation itself. In addition, removing Stage~I alignment lowers F1 to 73.42\%, confirming that stable cross-modal correspondence is a prerequisite for effective Fisher-guided fusion.

\textbf{Verification of Structural Necessity.}
Graph perturbation experiments further show that the model depends on the true topology of CPG. Randomly rewiring 90\% of the edges causes F1 to drop by 20.9\%, indicating that the model captures genuine structural information rather than merely exploiting node features. Additional discussion on calibration behavior and the interpretation of these ablations is provided in Appendix~\ref{appendix:ablation_details}.

\vspace{-0.4cm}

\begin{table*}[htbp]
\centering
\scriptsize
\setlength{\tabcolsep}{2pt}
  \caption{Ablation study results of TaCCS-DFA on the BigVul dataset. All experiments use CodeBERT as the backbone network.}
  \vspace{-0.3cm}
  \label{tab:ablation}
  \resizebox{\linewidth}{!}{%
  \begin{tabular}{@{}lcccccc@{}}
  \toprule
  \textbf{Experimental Setting} & Precision & Recall & ACC & F1 & ECE $\downarrow$ & $\Delta$F1 \\
  \midrule
  \multicolumn{7}{@{}l}{\cellcolor[HTML]{9ECFD4}{\textit{Core Component Ablation}}} \\
  TaCCS-DFA (Full Model) & \textbf{0.9667} & \textbf{0.6744} & \textbf{0.9716} & \textbf{0.7945} & \textbf{0.0163} & --- \\
  \quad w/o Fisher Guidance (Standard Attention) & 0.9655 & 0.6512 & 0.9697 & 0.7778 & 0.0295 & -2.1\% \\
  \quad w/ Random Fisher Bases $B_{\text{rand}}$ & 0.8485 & 0.6512 & 0.9621 & 0.7368 & 0.0231 & -7.3\% \\
  \quad w/ Slow Fisher Updates ($\text{freq}=2400$) & 0.9333 & 0.6512 & 0.9678 & 0.7671 & 0.0255 & -3.4\% \\
  \quad w/o InfoNCE Alignment & 0.8421 & 0.7442 & 0.9678 & 0.7901 & 0.0341 & -0.6\% \\
  \quad w/o Adaptive Gating (Fixed Fusion) & 0.9032 & 0.6512 & 0.9659 & 0.7568 & 0.0249 & -4.7\% \\
  \midrule
  \multicolumn{7}{@{}l}{\cellcolor[HTML]{9ECFD4}{\textit{Fisher Estimation Method Ablation}}} \\
  TaCCS-DFA (Oja, Default) & 0.9667 & 0.6744 & 0.9716 & 0.7945 & 0.0163 & --- \\
  \quad w/ Direct SVD & 0.9032 & 0.6512 & 0.9659 & 0.7568 & 0.0262 & -4.7\% \\
  \quad w/ Power Iteration & 0.8438 & 0.6279 & 0.9602 & 0.7200 & 0.0237 & -9.4\% \\
  \quad w/ Randomized SVD & 0.8485 & 0.6512 & 0.9621 & 0.7368 & 0.0260 & -7.3\% \\
  \quad w/ Batch SVD (No EMA) & 1.0000 & 0.6047 & 0.9678 & 0.7536 & 0.0279 & -5.1\% \\
  \midrule
  \multicolumn{7}{@{}l}{\cellcolor[HTML]{9ECFD4}{\textit{Verification of Structural Necessity}}} \\
  TaCCS-DFA (Full Model) & 0.9667 & 0.6744 & 0.9716 & 0.7945 & 0.0163 & --- \\
  \quad w/o Stage1 Alignment & 0.8056 & 0.6744 & 0.9602 & 0.7342 & 0.0276 & -7.6\% \\
  \quad w/ Edge Shuffle (90\% rewired) & 0.8148 & 0.5116 & 0.9508 & 0.6286 & 0.0407 & -20.9\% \\
  \quad w/ Degree-Preserving Rewire & 0.8750 & 0.6512 & 0.9640 & 0.7467 & 0.0301 & -6.0\% \\
  \quad w/ Remove DDG edges & 0.8788 & 0.6744 & 0.9659 & 0.7632 & 0.0264 & -3.9\% \\
  \quad w/ Remove CDG edges & 0.8529 & 0.6744 & 0.9640 & 0.7532 & 0.0241 & -5.2\% \\
  \midrule
  \multicolumn{7}{@{}l}{\cellcolor[HTML]{9ECFD4}{\textit{Theoretical Verification}}} \\
  Fisher Subspace Energy Ratio & --- & --- & --- & --- & 76.7\% & --- \\
  \quad vs. Random Orthogonal Baseline & --- & --- & --- & --- & +19.2\% & --- \\
  Adaptive Gating Retention ($1-\rho$) & --- & --- & --- & --- & 36.2\% & --- \\
  \bottomrule
  \end{tabular}
  }
\vspace{-0.2cm}
\end{table*}

\vspace{-0.3cm}

\subsection{RQ3: Interpretability and Mechanism Analysis}
\begin{figure*}[t]
\centering
\includegraphics[width=0.9\linewidth]{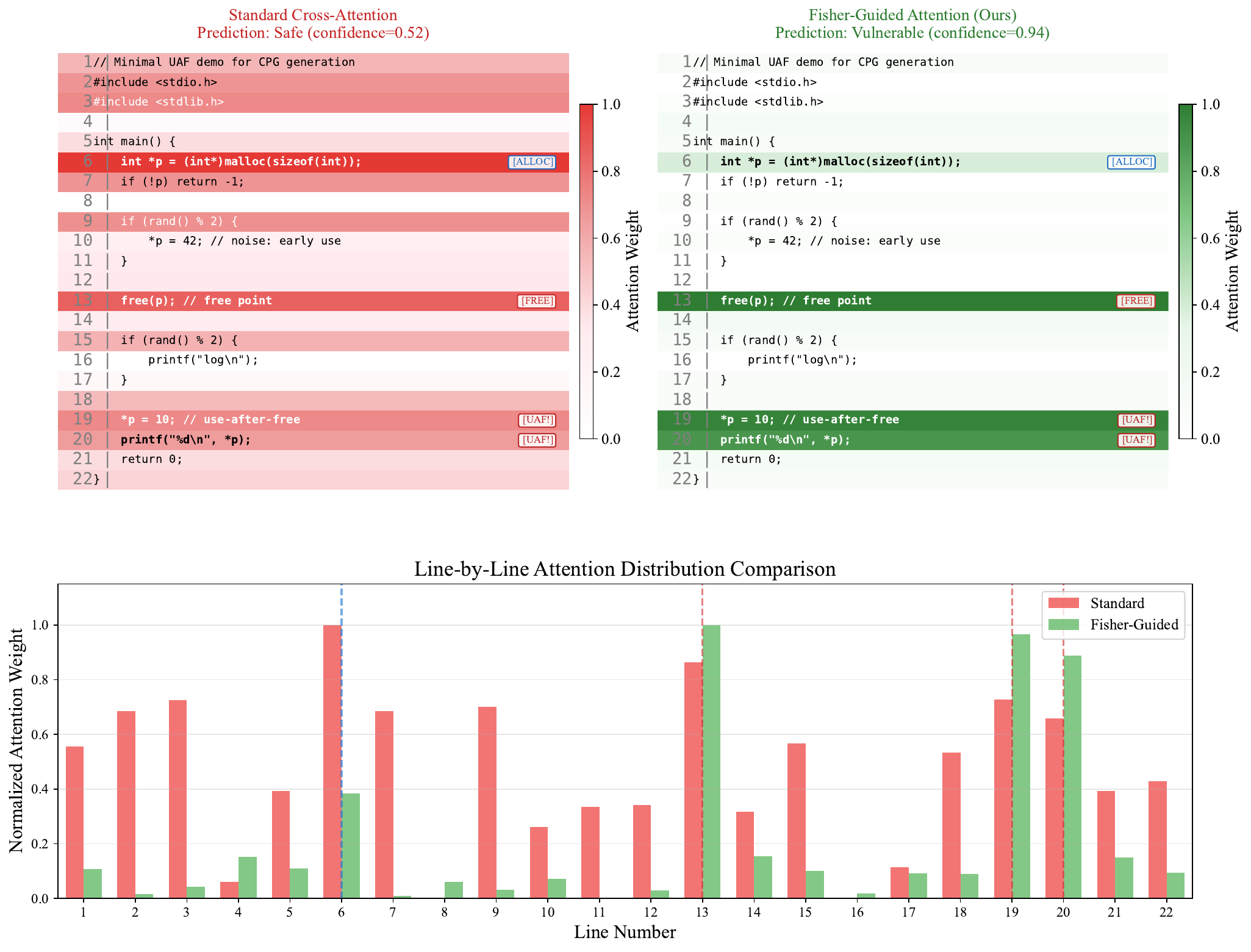}
\caption{
Line-level attention on a CWE-416 UAF sample. Standard cross-attention (left) diffuses over irrelevant lines and misclassifies the sample, whereas Fisher-guided attention (right) concentrates on the causal path---allocation, free, and subsequent use-after-free---and detects the vulnerability correctly. The bottom bar chart shows the per-line weight changes, with red dashed lines marking the key lines.
}
\vspace{-0.6cm}
\label{fig:case_study}
\end{figure*}

As shown in Figure~\ref{fig:case_study}, we compare the attention distribution on a CWE-416 (UAF) sample. Under the standard attention pattern, weights are diffusely distributed across irrelevant code lines, leading to erroneous prediction. In contrast, under the Fisher-guided pattern, attention precisely focuses on the three key points---memory allocation (malloc), release (free), and illegal access---forming a complete UAF vulnerability chain. The weights of the key lines increase by 170\%--200\%, correctly detecting the vulnerability with a confidence of 0.94.

Additional quantitative evidence on Fisher subspace energy concentration, the noise sensitivity experiment, and failure case analysis is provided in Appendix~\ref{appendix:mechanism_details}.

\subsection{RQ4: Efficiency and Scalability}
After introducing complex second-order information, computational efficiency becomes a key indicator for measuring the practicality of the method. Table~\ref{tab:efficiency} compares the performance of each model in terms of parameter count, training time, inference latency, and GPU memory usage, thereby evaluating the practical deployment potential of TaCCS-DFA.

Experimental results show that the per-batch training time of TaCCS-DFA is 2.29 seconds, slightly higher than the 2.27 seconds of Cross-Attention (+0.9\%). Although the Oja algorithm introduces additional gradient projection computations, the overall training overhead remains at the same order of magnitude as the complex baselines. Compared with the SOTA method Vul-LMGNN (2.42 sec/batch), TaCCS-DFA still maintains a competitive advantage.

Inference latency is another key metric. Although the Fisher projection matrix $\mathbf{U}$ introduces additional computation steps, the per-sample inference time of TaCCS-DFA is 22.27 ms, on the same order of magnitude as the 21.53 ms of the standard attention model, achieving a 10.0\% improvement in F1 score with only a 3.4\% increase in latency cost. Additionally, GPU memory usage maintains stability at 19.65 GB, representing a 0.1\% reduction compared to the baseline, indicating that the incremental PCA algorithm effectively controls the memory peak and enables the model to be deployed on large-scale code repositories.

TaCCS-DFA achieves a significant improvement in detection performance while maintaining a balance between computational efficiency and resource consumption, demonstrating strong engineering scalability. A visual comparison across efficiency metrics is provided in Appendix~\ref{appendix:efficiency_triptych}.
\begin{table}[t]
\centering
\footnotesize
\setlength{\tabcolsep}{2.5pt}
\caption{Comparison of computational efficiency and resource consumption. All experiments are conducted on the BigVul dataset with a batch size of 64.}
\label{tab:efficiency}
\begin{tabular}{@{}lccccc@{}}
\toprule
\multirow{2}{*}{\textbf{Model}} & \textbf{Params} & \textbf{Training Time} & \textbf{Inference} & \textbf{GPU Mem.} & \multirow{2}{*}{\textbf{F1}} \\
& (M) & (sec/batch) & (ms/sample) & (GB) & \\
\midrule
CodeBERT (NCS) & 125 & 2.28 & 13.6 & 19.7 & 0.6364 \\
RGCN (CPG) & 0.2 & 0.13 & 7.7 & 0.03 & 0.1850 \\
\midrule
ConcatFusion & 125 & 3.13 & 21.5 & 19.8 & 0.6944 \\
Cross-Attention & 129.77 & 2.27 & 21.53 & 19.67 & 0.7222 \\
Vul-LMGNN (C-B) & 125.2 & 2.42 & 14.7 & 20.8 & 0.7436 \\
\midrule
\textbf{TaCCS-DFA} & 127.85 & 2.29 & \textbf{22.27} & 19.65 & \textbf{0.7945} \\
\quad \textit{vs. Cross-Attn} & \textit{$-$1.5\%} & \textit{+0.9\%} & \textit{+3.4\%} & \textit{$-$0.1\%} & \textit{\textbf{+10.0\%}} \\
\bottomrule
\end{tabular}
\vspace{-0.3cm}
\end{table}

\section{Related Work}
\subsection{Deep Learning for Vulnerability Detection}
 Unimodal methods can be divided into two categories: sequence-based models leverage pretrained language models (such as CodeBERT \cite{feng2020codebert} and CodeT5 \cite{wang2021codet5}) or large language models (LLMs) \cite{ding2024vulnerability} to capture code semantics, but often lack explicit modeling of data flow and control flow, making them prone to over-reliance on spuriously correlated features such as variable names \cite{liu2025vul}; graph-based models (such as Devign \cite{zhou2019devign} and SySeVR \cite{li2021sysevr}) utilize graph neural networks to encode program structure, but are limited by the expressive power of graph encoders and the lack of semantic information in node initialization, resulting in discriminative performance that is typically weaker than that of pretrained sequence models.

To address the limitations of unimodal approaches, sequence-graph joint modeling has become a recent research hotspot. Representative works include GraphCodeBERT \cite{guo2020graphcodebert}, which uses data flow graphs to guide attention, and Vul-LMGNNs \cite{liu2025vul}, which employs language models to initialize graph nodes. However, these methods typically carry a strong implicit assumption that the introduction of an additional modality will necessarily bring effective information gains. In practice, because pretrained models have already implicitly encoded substantial structural information, significant feature redundancy exists between modalities. Meanwhile, the asymmetry in feature extraction capabilities between graph encoders and language models makes it difficult for simple concatenation or attention mechanisms to extract complementary information, often leading to the introduction of noise.

\subsection{Multimodal Fusion Mechanisms in SE}
In the field of software engineering, multimodal fusion mechanisms have mainly evolved from ``simple concatenation'' to ``attention-based interaction''. Early works adopted feature concatenation strategies \cite{wang2016automatically,russell2018automated}, directly concatenating feature vectors from different modalities before feeding them into a classifier. While simple, this approach cannot handle nonlinear interactions between modalities and is highly susceptible to interference from low-quality modal features.

Recent studies tend to use cross-attention mechanisms to dynamically aggregate information \cite{wan2019multi,tao2023vulnerability}. Although these methods are more flexible than concatenation, existing attention mechanisms primarily compute weights based on local similarity between features. This content-driven interaction has inherent limitations: it cannot distinguish which feature directions are truly important for the final classification task. When faced with highly redundant or noisy graph modalities, similarity-based attention may overly focus on redundant information that overlaps semantically with the primary modality, while overlooking the truly complementary subspaces that can improve the decision boundary.

\subsection{Fisher Information in Deep Learning}
The Fisher Information Matrix (FIM) is a core concept in information geometry, used to quantify the sensitivity of model outputs to changes in parameters. In deep learning, FIM has been widely applied to natural gradient optimization \cite{amari1998natural}, mitigation of catastrophic forgetting in continual learning (e.g., EWC \cite{kirkpatrick2017overcoming}), and uncertainty estimation \cite{ritter2018scalable}. Most of these works focus on the application of FIM in the parameter space, viewing it as a preconditioner for improving optimization paths or a constraint to prevent critical parameter drift.

The fundamental distinction of this paper's work from existing studies lies in the shift of application perspective: we extend FIM from the parameter space to the feature space, using it as a geometric metric for measuring ``task relevance''. Unlike traditional content-similarity-based fusion, this paper utilizes the principal subspace structure of FIM to identify feature directions that are most sensitive to classification loss, thereby guiding the cross-modal attention mechanism to filter redundant noise and retain only those structural features that make substantial contributions to the vulnerability detection task.

\section{Threats to Validity}
\textbf{Internal Validity.}
The Fisher principal subspace is approximated through online Oja updates, which may be affected by gradient noise and non-stationarity during training. To address this, we delay the initiation of Fisher estimation until cross-modal alignment becomes stable, and combine periodic updates, momentum smoothing, and orthogonalization to enhance estimation stability. The ablation experiments in Table~\ref{tab:ablation} and the supplementary mechanism analyses in Appendix~\ref{appendix:mechanism_details} validate the effectiveness of this approximation strategy.

\textbf{External Validity.}
We evaluate on four public datasets with markedly different vulnerability ratios, which reduces but does not eliminate the risk that the conclusions are dataset-specific. However, the datasets are still dominated by C/C++ code, and the method depends on Joern-generated CPGs; parsing failures or incomplete graphs may therefore affect performance. Although the framework is largely language-agnostic and Joern supports languages such as Java, JavaScript, and Python, we have not yet validated multi-language or inter-procedural vulnerability detection.


\textbf{Construct Validity.}
This paper adopts P/R/Acc/F1 and ECE as evaluation metrics, but does not take into account the manual review cost incurred by false positives in actual security auditing. This can be regarded as an extension direction for future work. To reduce experimental comparison bias, we reproduced the main baseline methods under unified data splits and training configurations, and designed control experiments such as parameter count alignment (Table~\ref{tab:ablation}); for some published works, we cite the numerical values reported in their original papers as references.

\section{Conclusion}
This paper studies the multimodal fusion dilemma in code vulnerability detection and shows that simple concatenation or generic cross-attention may dilute the discriminative signals of the dominant modality under modal redundancy and asymmetric feature extraction capability. To address this issue, we propose \mymethod, which combines task-conditioned feature selection, Fisher-guided cross-modal interaction, and adaptive gating to selectively extract complementary structural features from CPG while preserving the semantic advantages of NCS. Theoretically, DFA yields a tighter perturbation upper bound than full-spectrum attention under isotropic noise, and experiments on four benchmark datasets show consistent improvements across multiple backbones with good calibration and acceptable overhead.

Compared with LLMs, TaCCS-DFA is better suited as a low-latency, interpretable structural filtering module; few-shot LLM baselines remain substantially below it on function-level vulnerability detection (Appendix~\ref{appendix:llm_results}).

Future work includes extending the framework to other languages and finer-grained settings, exploring stronger graph encoders, and relaxing the isotropic noise assumption to analyze robustness under more general perturbation models.

\begin{credits}
\subsubsection{\ackname} 
This study was funded by the “Light of the West” Talent Training Program of Chinese Academy of Sciences (grant number XXXXXXXX).

\subsubsection{\discintname}
The authors have no competing interests to declare that are relevant to the content of this article.
\end{credits}

\bibliographystyle{splncs04}
\bibliography{references}

\newpage
\appendix
\noindent The appendices are ordered by the first citation order in the main text. Each appendix therefore appears immediately after the point where its supporting role is introduced in the paper narrative.

\section{Comparison with Traditional Static Analyzers}
\label{appendix:static_analysis}
This appendix supplements the motivation discussion in Section~2 by situating TaCCS-DFA against widely used rule-based security tools. It clarifies that our claim is not that CPG is harmful, but that learning-based structural modeling is necessary to go beyond the limited coverage of hand-crafted rules.

Traditional static analyzers rely on hand-crafted rules or pattern matching, offering the advantages of requiring no training data and enabling direct deployment; they are widely adopted foundational security tools in industry. To clearly quantify the performance gap between TaCCS-DFA and such tools as well as their complementary positioning, we systematically evaluated three mainstream static analyzers---Semgrep, Cppcheck, and Flawfinder---on the four benchmark datasets. Semgrep used the official \texttt{p/security-audit} rule set, Cppcheck was run with the \texttt{--enable=warning,performance,portability} options, and Flawfinder used its built-in rule library; all tools were executed with default configurations. The evaluation metrics were kept consistent with the main experiments.

The experimental results are presented in Table~\ref{tab:static_analysis_comparison}. The F1 scores of the three static analyzers across all four datasets are below 25\%, exhibiting a double disadvantage of both low precision and low recall. Taking BigVul as an example, the F1 scores of Semgrep, Cppcheck, and Flawfinder are only 2.43\%, 6.14\%, and 9.06\%, respectively, whereas TaCCS-DFA achieves 87.80\% on the same dataset. On the Devign dataset where positive and negative samples are approximately balanced, Semgrep's recall is merely 0.57\%, indicating that rule-based methods suffer from severely insufficient generalization when confronted with diverse real-world vulnerability patterns.

Traditional static analyzers are inherently limited by the coverage of their rules, resulting in orders-of-magnitude gaps in detection performance compared with deep learning methods. TaCCS-DFA, by fusing code semantics and program structure information, is able to capture complex vulnerability patterns that are difficult to describe with static rules. The two categories of methods are complementary in application scenarios: static analyzers are suitable as lightweight rapid screening tools, while TaCCS-DFA is better suited for in-depth semantic and structural joint analysis of high-risk code.

\begin{table*}[htbp]
\centering
\scriptsize
\setlength{\tabcolsep}{2pt}
\caption{Comparison of Vulnerability Detection Performance between Traditional Static Analyzers and \mymethod}
\label{tab:static_analysis_comparison}
\sisetup{
  table-format=2.2,
  round-mode=places,
  round-precision=2,
  detect-weight,
  detect-display-math=true
}
\resizebox{\linewidth}{!}{%
\begin{tabular}{@{}l *{4}{S[table-format=2.2]} *{4}{S[table-format=2.2]} *{4}{S[table-format=2.2]} *{4}{S[table-format=2.2]}@{}}
\toprule
\textbf{Model}
& \multicolumn{4}{c}{\textbf{BigVul}}
& \multicolumn{4}{c}{\textbf{PrimeVul}}
& \multicolumn{4}{c}{\textbf{Devign}}
& \multicolumn{4}{c}{\textbf{ReVeal}} \\
\cmidrule(lr){2-5} \cmidrule(lr){6-9} \cmidrule(lr){10-13} \cmidrule(lr){14-17}
& {P} & {R} & {ACC} & {F1}
& {P} & {R} & {ACC} & {F1}
& {P} & {R} & {ACC} & {F1}
& {P} & {R} & {ACC} & {F1} \\
\midrule
\multicolumn{17}{l}{\cellcolor[HTML]{9ECFD4}{\textit{Traditional Static Analyzers}}} \\
\Xhline{0.5pt}
Semgrep (official) & 7.21 & 1.46 & 96.36 & 2.43 & 9.14 & 3.10 & 97.17 & 4.63 & 31.82 & 0.57 & 53.71 & 1.12 & 7.69 & 0.46 & 89.65 & 0.87 \\
Cppcheck & 4.99 & 7.99 & 92.42 & 6.14 & 3.49 & 11.84 & 90.79 & 5.39 & 54.55 & 6.35 & 54.49 & 11.37 & 19.51 & 11.06 & 86.75 & 14.12 \\
Flawfinder & 6.34 & 15.89 & 90.10 & 9.06 & 5.75 & 27.69 & 88.35 & 9.52 & 49.44 & 14.40 & 53.85 & 22.31 & 16.20 & 21.20 & 81.43 & 18.36 \\
\midrule
\multicolumn{17}{l}{\cellcolor[HTML]{9ECFD4}{\textit{Ours}}} \\
\Xhline{0.5pt}
\textbf{\mymethod (CodeBERT)} & \textbf{96.67} & \textbf{67.44} & \textbf{97.16} & \textbf{79.45} & \textbf{22.07} & \textbf{29.51} & \textbf{96.13} & \textbf{25.25} & \textbf{59.54} & \textbf{65.44} & \textbf{64.30} & \textbf{62.35} & \textbf{51.43} & \textbf{53.73} & \textbf{89.96} & \textbf{52.55} \\
\textbf{\mymethod (CodeT5-Small)} & \textbf{94.44} & \textbf{79.07} & \textbf{97.92} & \textbf{86.08} & \textbf{17.93} & \textbf{29.33} & \textbf{95.46} & \textbf{22.25} & \textbf{57.03} & \textbf{70.44} & \textbf{62.00} & \textbf{63.03} & \textbf{39.34} & \textbf{70.65} & \textbf{85.69} & \textbf{50.53} \\
\textbf{\mymethod (CodeT5-Base)} & \textbf{92.31} & \textbf{83.72} & \textbf{98.11} & \textbf{87.80} & \textbf{23.50} & \textbf{23.86} & \textbf{96.60} & \textbf{23.68} & \textbf{57.40} & \textbf{73.86} & \textbf{62.77} & \textbf{64.60} & \textbf{51.60} & \textbf{56.22} & \textbf{90.02} & \textbf{53.81} \\
\bottomrule
\end{tabular}%
}
\flushleft{\small Semgrep uses the official \texttt{p/security-audit} rule set; Cppcheck uses \texttt{--enable=warning,performance,portability}; Flawfinder uses its built-in rules. }
\end{table*}

\section{Proof of Theorem}
\label{appendix:proof}
This appendix provides the proof sketch referenced at the end of the theoretical robustness analysis. The goal is to make explicit where the Lipschitz assumption enters and how the projection onto the Fisher principal subspace yields the $\sqrt{k/d}$ contraction term.
\begin{proof}[Proof sketch.]
To prove Theorem~\ref{thm:bound}, we first present a unified upper-bound form for full-spectrum attention and DFA under input perturbations. Let
\(\tilde{\mathbf{H}}_{\text{cpg}} = \mathbf{H}_{\text{cpg}} + \bm{\Delta}\), with
\(\|\bm{\Delta}\|_F \le \varepsilon\).
Assume that the attention operator \(\mathcal{F}(\cdot)\) is
\(L\)-Lipschitz continuous with respect to the input \(\mathbf{H}_{\text{cpg}}\) in the Frobenius norm:
\begin{equation}
\|\mathcal{F}(\mathbf{H}_1) - \mathcal{F}(\mathbf{H}_2)\|_F
\le L \cdot \|\mathbf{H}_1 - \mathbf{H}_2\|_F .
\label{eq:lipschitz}
\end{equation}
\textbf{Perturbation Upper Bound for Full-Spectrum Attention}
For the full-spectrum attention mechanism \(\mathcal{F}_{\text{full}}\), its input is directly
\(\mathbf{H}_{\text{cpg}}\). From Equation~\eqref{eq:lipschitz} we obtain
\begin{equation}
\|\mathcal{F}_{\text{full}}(\tilde{\mathbf{H}}_{\text{cpg}})
 - \mathcal{F}_{\text{full}}(\mathbf{H}_{\text{cpg}})\|_F
\le L \cdot \|\tilde{\mathbf{H}}_{\text{cpg}} - \mathbf{H}_{\text{cpg}}\|_F
= L \cdot \|\bm{\Delta}\|_F
\le L \cdot \varepsilon,
\end{equation}
which yields the upper bound for \(\mathcal{F}_{\text{full}}\) stated in the theorem.
\textbf{Effective Input Perturbation of DFA}
For TaCCS-DFA, we employ the orthogonal projection matrix
\(\mathbf{P} = \mathbf{U}\mathbf{U}^\top \in \mathbb{R}^{d \times d}\) in Complementary Subspace Attention to restrict the auxiliary modality representations to the Fisher principal subspace
\(\mathcal{S}_{\text{fisher}} = \mathrm{span}(\mathbf{U})\), and construct the key and value vectors based on \(\mathbf{H}_{\text{cpg}}\mathbf{P}\).
Under the noise decomposition
\(\bm{\Delta} = \bm{\Delta}_\parallel + \bm{\Delta}_\perp\), where
\(\bm{\Delta}_\parallel = \bm{\Delta}\mathbf{P}\),
\(\bm{\Delta}_\parallel\) exactly corresponds to the component of the noise lying in the Fisher principal subspace.
Equivalently, the cross-modal attention of DFA can be expressed as an operator that depends only on \(\mathbf{H}_{\text{cpg}}\mathbf{P}\):
\begin{equation}
\mathcal{F}_{\text{dfa}}(\mathbf{H}_{\text{cpg}})
= \mathcal{F}_{\text{full}}(\mathbf{H}_{\text{cpg}}\mathbf{P}),
\end{equation}
so that after perturbation we have
\begin{equation}
\mathcal{F}_{\text{dfa}}(\tilde{\mathbf{H}}_{\text{cpg}})
= \mathcal{F}_{\text{full}}((\mathbf{H}_{\text{cpg}}+\bm{\Delta})\mathbf{P})
= \mathcal{F}_{\text{full}}(\mathbf{H}_{\text{cpg}}\mathbf{P}
+ \bm{\Delta}\mathbf{P}).
\end{equation}
Denoting \(\bm{\Delta}_\parallel = \bm{\Delta}\mathbf{P}\), the Lipschitz property yields
\begin{equation}
\begin{aligned}
\|\mathcal{F}_{\text{dfa}}(\tilde{\mathbf{H}}_{\text{cpg}})
 - \mathcal{F}_{\text{dfa}}(\mathbf{H}_{\text{cpg}})\|_F
&\;\lesssim\;
\|\mathcal{F}_{\text{full}}(\mathbf{H}_{\text{cpg}}\mathbf{P}
 + \bm{\Delta}_\parallel) \\
&\quad - \mathcal{F}_{\text{full}}(\mathbf{H}_{\text{cpg}}\mathbf{P})\|_F \\
&\le L \cdot \|\bm{\Delta}_\parallel\|_F
 + o(\|\bm{\Delta}\|_F),
\end{aligned}
\label{eq:dfa-lipschitz}
\end{equation}
where \(o(\|\bm{\Delta}\|_F)\) absorbs the higher-order terms of nonlinear operations such as Softmax under small perturbations.
\textbf{Energy Contraction under Isotropic Noise}
We next exploit the isotropic noise assumption: given \(\|\bm{\Delta}\|_F\), the direction of \(\bm{\Delta}\) is uniformly distributed over the \(d\)-dimensional feature space, so its energy is evenly distributed across all dimensions. Consequently, the expected proportion of noise energy falling into any \(k\)-dimensional orthogonal subspace is \(k/d\). Formally,
\begin{equation}
\mathbb{E}\big[\|\bm{\Delta}_\parallel\|_F^2\big]
= \mathbb{E}\big[\|\bm{\Delta}\mathbf{P}\|_F^2\big]
= \frac{k}{d} \|\bm{\Delta}\|_F^2
\le \frac{k}{d} \varepsilon^2.
\end{equation}
By Jensen's inequality
\(\mathbb{E}\|X\| \le \sqrt{\mathbb{E}\|X\|^2}\) we obtain
\begin{equation}
\mathbb{E}\big[\|\bm{\Delta}_\parallel\|_F\big]
\le \sqrt{\mathbb{E}\big[\|\bm{\Delta}_\parallel\|_F^2\big]}
\le \sqrt{\frac{k}{d}} \, \varepsilon.
\label{eq:proj-bound}
\end{equation}
\textbf{Overall Expected Perturbation Upper Bound for DFA}
Substituting Equation~\eqref{eq:proj-bound} into both sides of the expectation of Equation~\eqref{eq:dfa-lipschitz} gives
\begin{equation}
\mathbb{E}\left[
\|\mathcal{F}_{\text{dfa}}(\tilde{\mathbf{H}}_{\text{cpg}})
 - \mathcal{F}_{\text{dfa}}(\mathbf{H}_{\text{cpg}})\|_F
\right]
\le L \cdot \mathbb{E}\big[\|\bm{\Delta}_\parallel\|_F\big]
+ o(\varepsilon)
\le L \cdot \sqrt{\frac{k}{d}} \cdot \varepsilon + o(\varepsilon),
\end{equation}
which is precisely the DFA perturbation upper bound given in Theorem~\ref{thm:bound}.
Since \(k \ll d\), the noise suppression factor \(\sqrt{k/d}\) is significantly smaller than 1, demonstrating that TaCCS-DFA theoretically possesses a tighter risk upper bound compared to full-spectrum attention.
\end{proof}

\section{Dataset CWE Distribution}
\label{appendix:dataset_cwe}
This appendix expands the dataset characterization referenced in Section~4.1. It clarifies which vulnerability families dominate the evaluation and why memory-safety vulnerabilities remain the primary focus of the reported benchmarks.

\begin{figure*}[t]
\centering
\includegraphics[width=\linewidth]{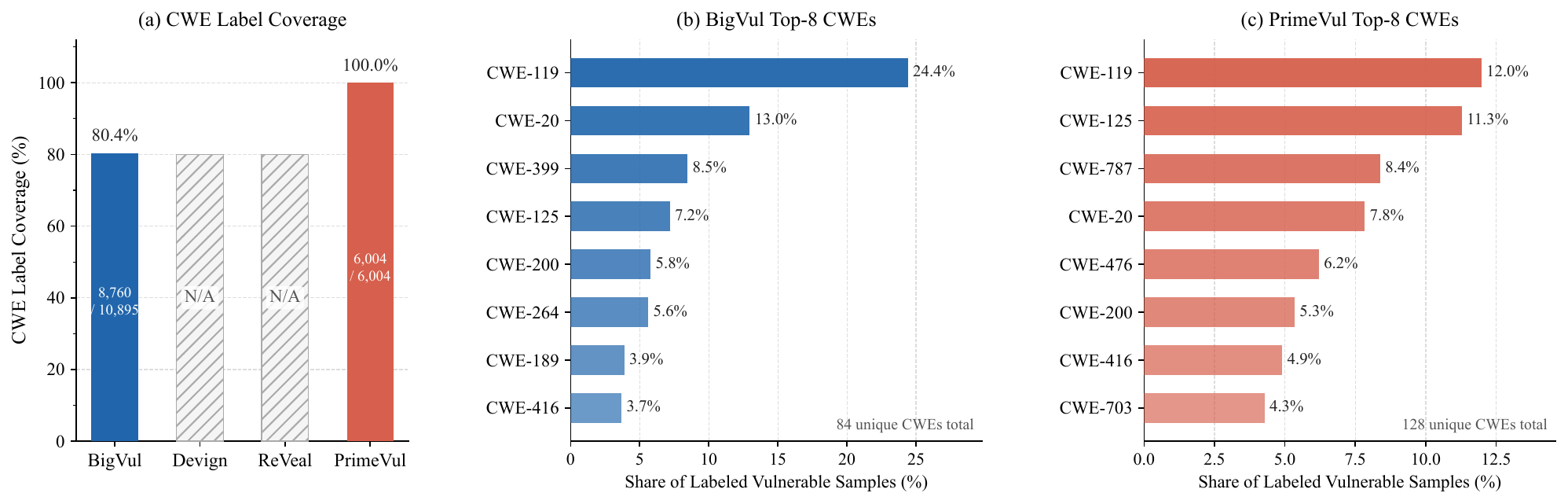}
\caption{Dataset CWE label coverage and vulnerability type distribution. (a) CWE annotation coverage rate of vulnerable samples in each dataset (Devign and ReVeal provide only binary labels marked as N/A); (b)(c) the eight most frequent CWE types in BigVul and PrimeVul and their proportions among the annotated vulnerable samples.}
\label{fig:dataset_cwe}
\end{figure*}

To further characterize the distribution of vulnerability types across the datasets, we perform statistical analysis on the standardized CWE labels of BigVul and PrimeVul; the results are shown in Figure~\ref{fig:dataset_cwe}. As illustrated in Figure (a), 80.4\% of the vulnerability samples in BigVul carry valid CWE annotations, while PrimeVul achieves 100\% coverage; Devign and ReVeal provide only binary labels. In terms of vulnerability type distribution, both datasets are dominated by memory-safety defects: CWE-119 (buffer error) accounts for 24.4\% in BigVul and 12.0\% in PrimeVul; CWE-125 (out-of-bounds read), CWE-20 (improper input validation), CWE-200 (information exposure), and CWE-416 (use-after-free) consistently appear among the Top-8 types in both datasets.

\section{Implementation Details}
\label{appendix:implementation_details}
This appendix collects the training and optimization details referenced in Section~4.1. We report the hyperparameters that are shared across all main experiments so that the Fisher-guided components can be reproduced without searching through code or scripts.

All experiments are conducted on four NVIDIA 3090 GPUs. Model training employs the AdamW optimizer for a total of 15 epochs. The global learning rate is set to $2\times10^{-5}$, the weight decay coefficient to 0.01, and the batch size is fixed at 64. In the configuration, the Fisher subspace dimension $k$ is set to 32. The momentum coefficient $\mu$ of the Oja algorithm is set to 0.99, and the projection matrix is updated every 1200 steps. The training process is divided into two stages to establish semantic correspondence between heterogeneous modalities. In the first stage, which covers the initial 48\% of the training epochs, the weight of the cross-modal alignment loss $\alpha$ is set to 0.05. In the subsequent fine-tuning stage, this weight is slightly reduced to 0.045. For the InfoNCE alignment loss, the temperature coefficient $\tau$ is set to 0.2.

\section{Detailed Few-Shot LLM Results}
\label{appendix:llm_results}
This appendix expands the brief comparison with general-purpose code LLMs in Section~4.2 and the conclusion. It reports the full metrics behind the statement that prompt-only inference still underperforms task-specific multimodal vulnerability detectors.

To compare task-specific multimodal models with general-purpose code LLMs, we further evaluate mainstream LLMs such as Qwen3-Coder and DeepSeek-V3 under few-shot settings. Overall, the LLM baselines exhibit a clear high-recall, low-precision imbalance and remain below TaCCS-DFA on both Devign and ReVeal, indicating that prompt-based learning alone cannot replace task-specific multimodal fusion for function-level vulnerability detection.
\begin{table*}[htbp]
\centering
\scriptsize
\setlength{\tabcolsep}{2pt}
\caption{Few-shot LLM results on the Devign and ReVeal datasets.}
\label{tab:llm_results}
\sisetup{
  table-format=2.2,
  round-mode=places,
  round-precision=2,
  detect-weight,
  detect-display-math=true
}
\resizebox{\linewidth}{!}{%
\begin{tabular}{@{}l *{4}{S[table-format=2.2]} *{4}{S[table-format=2.2]}@{}}
\toprule
\textbf{Model} & \multicolumn{4}{c}{\textbf{Devign}} & \multicolumn{4}{c}{\textbf{ReVeal}} \\
\cmidrule(lr){2-5} \cmidrule(lr){6-9}
& {P (\%)} & {R (\%)} & {ACC (\%)} & {F1 (\%)} & {P (\%)} & {R (\%)} & {ACC (\%)} & {F1 (\%)} \\
\midrule
Qwen3-Coder (480B) & 51.16 & 32.98 & 54.70 & 40.11 & 17.70 & 28.36 & 78.95 & 21.80 \\
DeepSeek-V3.1 (671B) & 48.88 & 47.98 & 53.01 & 48.43 & 19.24 & 37.81 & 77.15 & 25.50 \\
Ministral-3 (8B) & 47.57 & 83.33 & 50.10 & 60.57 & 10.90 & 78.61 & 31.34 & 19.15 \\
Gemma3 (27B) & 49.85 & 45.09 & 53.89 & 47.35 & 14.70 & 37.81 & 70.87 & 21.17 \\
GLM-4.6 & 48.64 & 62.54 & 52.20 & 54.72 & 13.69 & 50.75 & 61.81 & 21.56 \\
MiniMax-M2 & 46.62 & 48.95 & 50.75 & 47.75 & 10.90 & 78.61 & 31.34 & 19.15 \\
GPT-OSS (120B) & 49.43 & 49.21 & 53.49 & 49.32 & 11.33 & 54.73 & 51.00 & 18.77 \\
LLAMA-3.3 (70B) & 55.98 & 24.21 & 56.39 & 33.80 & 13.20 & 57.21 & 56.66 & 21.46 \\
\bottomrule
\end{tabular}%
}
\end{table*}

\section{Few-Shot LLM Evaluation Prompt Template}
\label{appendix:llm-fewshot-prompt}
This appendix provides the concrete prompt template corresponding to the few-shot LLM results above. We include it separately because the prompt design and fixed in-context examples directly affect the precision-recall trade-off of the evaluated LLM baselines.

Figure~\ref{fig:llm-fewshot-prompt} illustrates the prompt template used for few-shot LLM vulnerability detection evaluation. This template is built upon the zero-shot template by inserting $k=4$ examples (2 vulnerable samples + 2 non-vulnerable samples) after the system prompt and before the code under test. Each example consists of a user prompt (containing a code snippet) and an assistant response (JSON-formatted judgment result). The examples are randomly sampled from the training set and kept fixed throughout the entire testing process to prevent information leakage from the test set.
\begin{figure}[h]
\centering
\small
\fbox{\parbox{0.95\linewidth}{
\textbf{System Prompt:}\\[0.2em]
\texttt{You are a professional code security analyst. Your task is to analyze the given C code snippet and determine whether it contains security vulnerabilities.}\\[0.3em]
\texttt{Common vulnerability types include: Buffer overflow (CWE-119/120), Out-of-bounds read/write (CWE-125/787), Use-after-free (CWE-416), Integer overflow (CWE-190), NULL pointer dereference (CWE-476), Format string (CWE-134), Memory leak (CWE-401), Race condition (CWE-362).}\\[0.3em]
\texttt{You MUST respond in JSON format: \{"vulnerable": true/false, "confidence": 0.0-1.0\}}\\[0.8em]
\textbf{Example 1 (Vulnerable) - User:}\\[0.2em]
\texttt{Analyze the following C code for security vulnerabilities:}\\
\texttt{```c}\\
\texttt{void func(char *src) \{}\\
\texttt{~~char buf[10];}\\
\texttt{~~strcpy(buf, src); // Buffer overflow}\\
\texttt{\}}\\
\texttt{```}\\[0.2em]
\textbf{Example 1 - Assistant:}\\[0.2em]
\texttt{\{"vulnerable": true, "confidence": 0.9\}}\\[0.6em]
\textbf{Example 2 (Safe) - User:}\\[0.2em]
\texttt{Analyze the following C code for security vulnerabilities:}\\
\texttt{```c}\\
\texttt{int add(int a, int b) \{}\\
\texttt{~~return a + b;}\\
\texttt{\}}\\
\texttt{```}\\[0.2em]
\textbf{Example 2 - Assistant:}\\[0.2em]
\texttt{\{"vulnerable": false, "confidence": 0.95\}}\\[0.1em]
\textbf{......}\\[0.1em]
\textbf{Actual Query - User:}\\[0.2em]
\texttt{Analyze the following C code for security vulnerabilities:}\\
\texttt{```c}\\
\texttt{\{code\}}\\
\texttt{```}
}}
\caption{Prompt template used for few-shot LLM vulnerability detection evaluation. Four examples (2 vulnerable samples + 2 safe samples) are inserted after the system prompt. Each example consists of a user prompt (code snippet) and an assistant response (JSON judgment result). The examples are randomly sampled from the training set and kept fixed throughout the evaluation to prevent test set leakage.}
\label{fig:llm-fewshot-prompt}
\end{figure}

\section{Main Results Visualization}
\label{appendix:main_results_profile}
This appendix gives a compact visual summary of the four main metrics across datasets and model families. It is intended to make the precision-recall trade-off discussed in Section~4.2 easier to inspect at a glance.
\begin{figure*}[t]
\centering
\includegraphics[width=\linewidth]{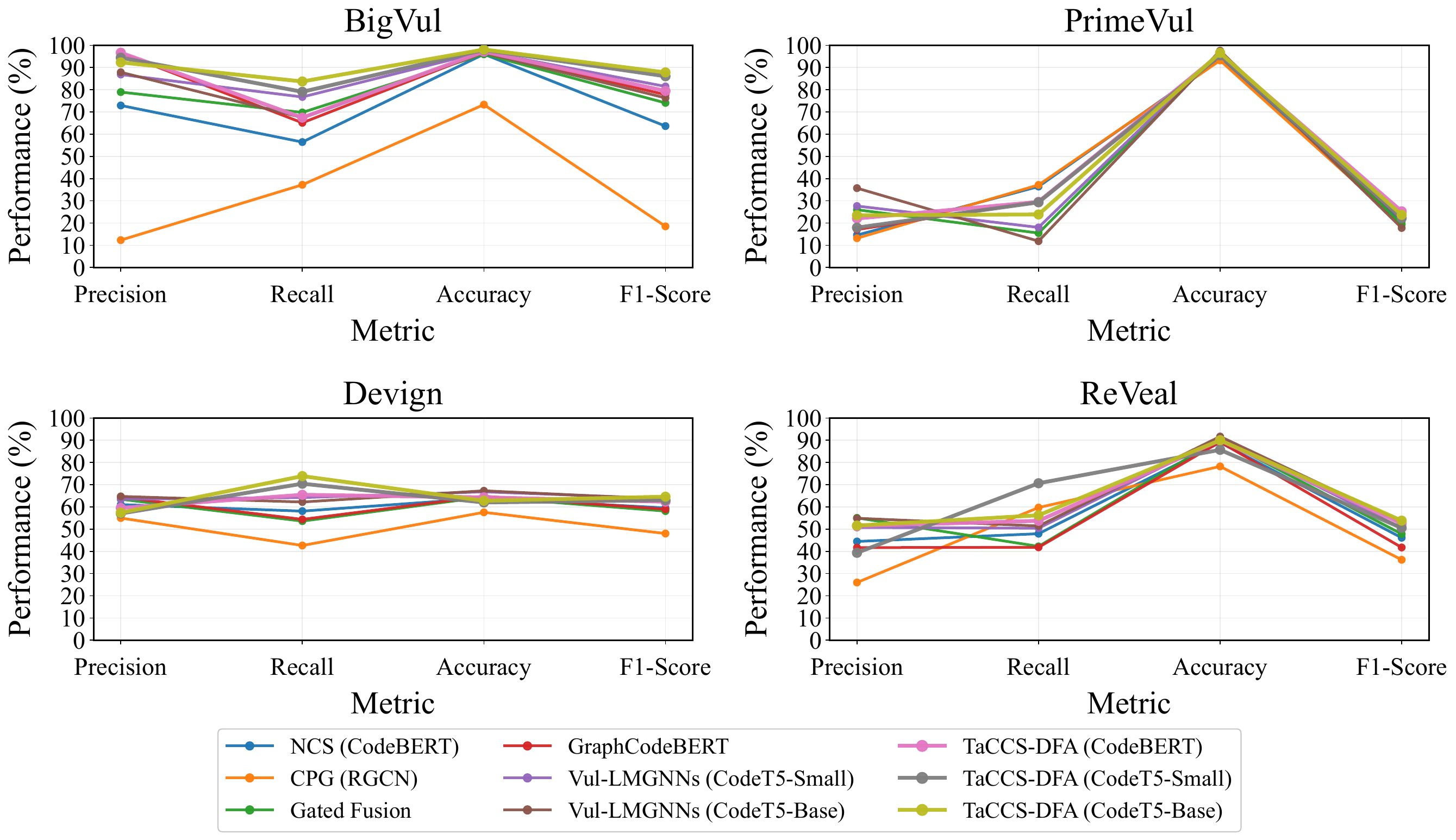}
\caption{Metrics profile of the main results on the four datasets. The horizontal axis represents Precision, Recall, Accuracy, and F1-Score, and the vertical axis represents the corresponding performance (\%). Each curve shows the four-metric performance of a model/method on that dataset.}
\vspace{-0.1cm}
\label{fig:main_results_profile}
\end{figure*}

\section{Extended Ablation Results}
\label{appendix:ablation_details}
This appendix expands the ablation discussion in Section~4.3 with additional interpretation of calibration, Fisher estimation alternatives, and structure-destruction controls. The purpose is to separate the effects of geometric guidance, parameter count, and true graph structure.

\textbf{Comparison of Fisher Estimation Methods}
Compared with alternative approaches such as Direct SVD ($O(d^3)$ complexity, difficult to scale to high dimensions), Power Iteration (slow convergence when the eigenvalue spectrum is flat), and Batch SVD (sensitive to batch fluctuations), Oja's Rule achieves a significantly superior F1 score of 79.45\% thanks to its linear $O(dk)$ complexity and implicit forgetting mechanism (naturally attenuating outdated gradient information through online updates). This enables it to smoothly adapt to dynamic changes during the training process.

\textbf{Model Calibration Analysis}
In addition to discriminative performance, Table~\ref{tab:ablation} reports the Expected Calibration Error (ECE) for each variant. The complete TaCCS-DFA model achieves the lowest ECE value, indicating that its output probabilities are not only discriminative but also faithfully reflect prediction confidence. In comparison, removing Fisher guidance increases ECE to 0.0295 (+81\%), while replacing the Fisher projection with a random orthogonal basis yields an ECE of 0.0231 (+42\%). These results demonstrate that Fisher subspace filtering effectively reduces the model's tendency toward overconfidence in high-confidence regions by suppressing the propagation of noisy features, thereby improving probability calibration.

\section{Supplementary Mechanism Analysis}
\label{appendix:mechanism_details}
This appendix provides the additional mechanism-level evidence referenced in Section~4.4 and the threats-to-validity discussion. It connects the empirical behavior of the model to the Fisher-subspace interpretation used throughout the paper.

\begin{figure}[t]
\centering
\includegraphics[width=0.75\linewidth]{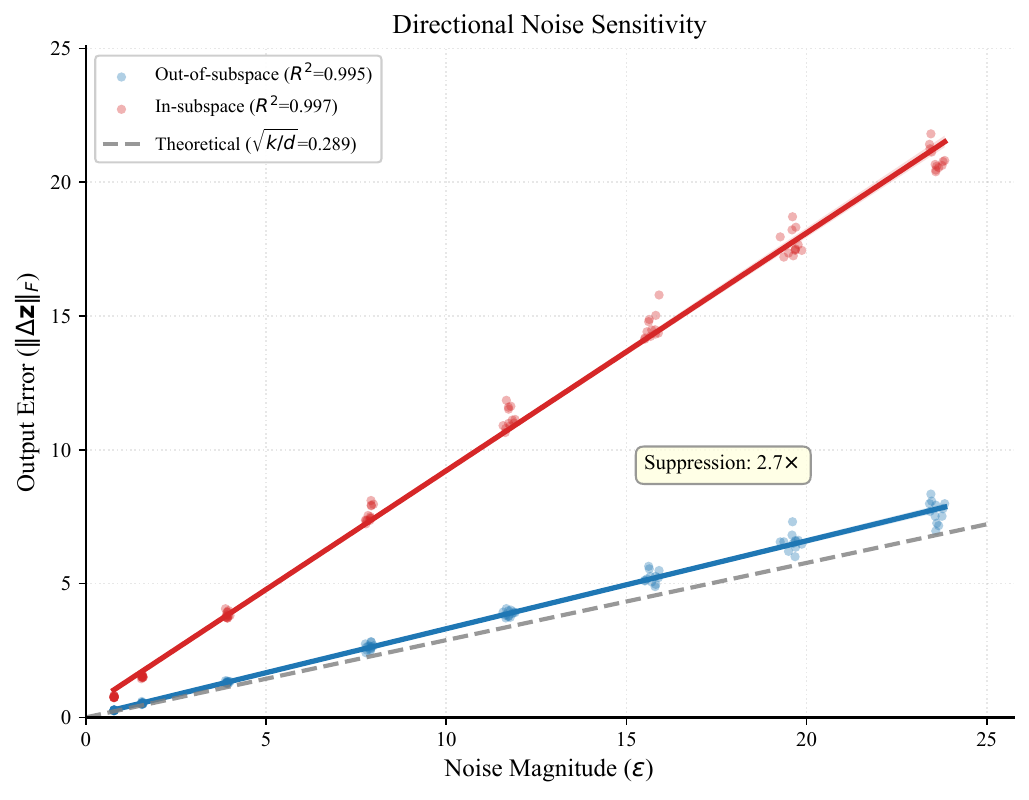}
\vspace{-0.3cm}
\caption{
Noise sensitivity experiment. The horizontal axis represents noise intensity $\varepsilon$, and the vertical axis represents output deviation $\|\Delta \mathbf{z}\|$. The red curve represents the case retaining only the orthogonal complement component ($\bm{\Delta}_{\perp}$), the blue curve represents the case retaining only the Fisher subspace component ($\bm{\Delta}_{\parallel}$), and the gray dashed line indicates the theoretical prediction of Theorem~\ref{thm:bound} under the isotropic noise assumption ($\sqrt{k/d}=0.289$).
}
\label{fig:noise_sensitivity}
\end{figure}

To quantitatively validate the Fisher guidance mechanism, we analyze the energy distribution of the DFA output. The Fisher Subspace Energy Ratio reaches 76.7\%, indicating that the majority of feature energy is concentrated on the task-sensitive low-dimensional manifold, representing a 19.2\% improvement compared to using a random orthogonal basis. The noise sensitivity experiment shown in Figure~\ref{fig:noise_sensitivity} further validates Theorem~\ref{thm:bound}: the orthogonal complement component ($\bm{\Delta}_{\perp}$) is filtered after projection and has almost no impact on the output; whereas the output deviation caused by the Fisher subspace component ($\bm{\Delta}_{\parallel}$) increases linearly with noise intensity, with a slope close to the theoretically predicted value $\sqrt{k/d}=0.289$.

\textbf{Failure Case Analysis}
Although TaCCS-DFA performs well in most scenarios, there are still two relatively weak cases. The first involves cross-procedure data-flow vulnerabilities, typically represented by inter-procedural use-after-free. Since CPG is constructed at the function-level granularity, it cannot capture data dependencies across function boundaries, resulting in incomplete vulnerability causal paths in the graph structure and significantly increased detection difficulty. The second involves vulnerability types where discriminative signals primarily derive from lexical patterns, such as format string vulnerabilities and integer overflows. In these cases, the adaptive gating mechanism automatically reduces the fusion weight of the graph modality, which is consistent with the low Adaptive Gating Retention reported in the ablation study.

\section{Efficiency Visualization}
\label{appendix:efficiency_triptych}
This appendix visualizes the efficiency comparison summarized in Section~4.5. It is intended to make the latency-memory-performance trade-offs more interpretable than the scalar table alone.
\begin{figure}[t]
\centering
\includegraphics[width=\linewidth]{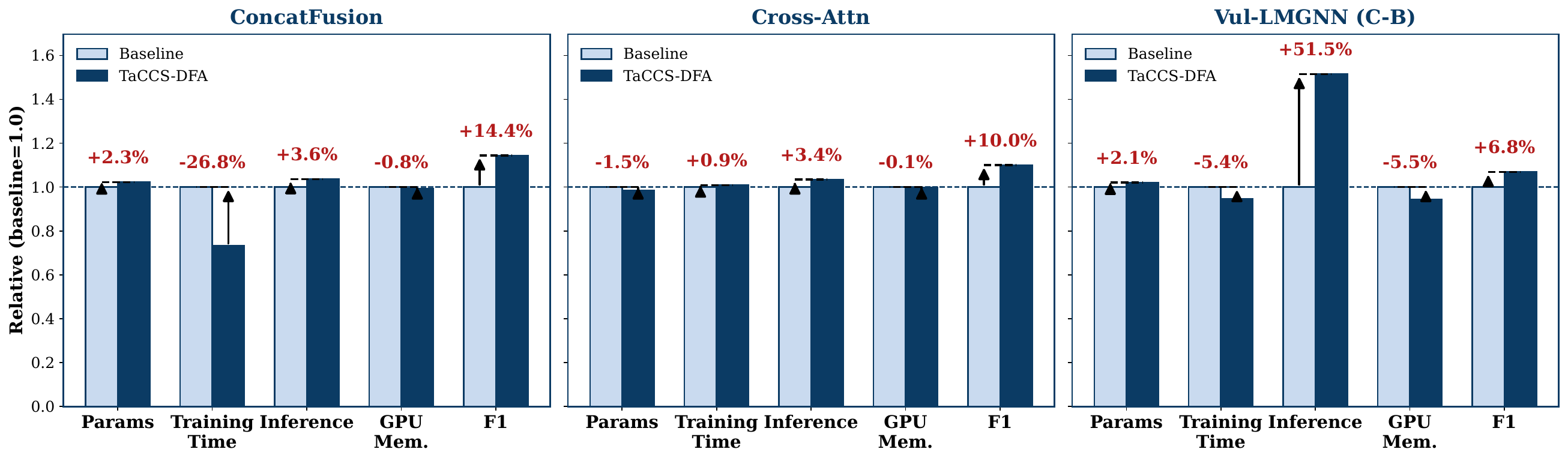}
\caption{Efficiency comparison of TaCCS-DFA with mainstream fusion methods. From left to right, the comparisons with ConcatFusion, Cross-Attention, and Vul-LMGNN (CodeBert) are shown, respectively; arrows and dashed lines highlight the relative changes in each metric.}
\label{fig:efficiency_triptych_appendix}
\vspace{-0.2cm}
\end{figure}

\end{document}